\documentclass[aps,prc,twocolumn,10pt,superscriptaddress]{revtex4-1}

\usepackage{hyperref}
\usepackage{graphicx}
\usepackage{amsmath}
\usepackage{slashed} 
\usepackage[dvipsnames,svgnames,table]{xcolor}
\usepackage[normalem]{ulem}  
\renewcommand\sout{\bgroup \color{red} \ULdepth=-.5ex \ULset}

\begin{document}

\title{Disentangling Initial-State and Evolution Effects in Heavy-Ion Collisions Using EPOS and PHSD}

\author{Mahbobeh Jafarpour}\email{mahbobeh.jafarpour@subatech.in2p3.fr}
\affiliation{SUBATECH, University of Nantes - IN2P3/CNRS - IMT Atlantique , Nantes, France}

\author{Vadym Voronyuk}
\affiliation{Joint Institute for Nuclear Research, Joliot-Curie 6, 141980 Dubna, Moscow region, Russia}

\author{Klaus Werner}\email{Werner@subatech.in2p3.fr}
\affiliation{SUBATECH, University of Nantes - IN2P3/CNRS - IMT Atlantique , Nantes, France}

\author{Elena Bratkovskaya}
\affiliation{GSI Helmholtzzentrum f\"ur Schwerionenforschung GmbH,Planckstrasse 1, D-64291 Darmstadt, Germany}
\affiliation{Institut f\"ur Theoretische Physik, Johann Wolfgang Goethe-Universit\"at,Max-von-Laue-Str.\ 1, D-60438 Frankfurt am Main, Germany}
\affiliation{Helmholtz Research Academy Hesse for FAIR (HFHF), GSI Helmholtz Center for heavy-ion Physics, Campus Frankfurt, 60438 Frankfurt, Germany}

\author{Damien Vintache} 
\affiliation{SUBATECH, University of Nantes - IN2P3/CNRS - IMT Atlantique , Nantes, France}

\begin{abstract}
In this study we examine the impact of the initial stage and dynamical evolution on final-state observables in heavy-ion collisions. For this goal we develop a novel approach, EPOSir+PHSDe, which employs EPOS initial conditions as the starting point for parton and hadron evolution within the PHSD microscopic transport approach. By examining the space-time evolution of matter in this model and comparing to EPOS (which starts with  an S-matrix approach for parallel scatterings for the initial conditions and uses a hydrodynamic evolution for the quark-gluon plasma stage with the UrQMD as afterburner) and PHSD (which starts with primary high energy $NN$ scattering realized via the LUND string model and continues with fully microscopic transport dynamics for strongly interacting partonic and hadronic matter), we identify the key differences in the final particle distributions among the three approaches. Our analysis focuses on rapidity, transverse momentum spectra, and flow harmonics $v_2$ for Au+Au collisions at the invariant energy of $\sqrt{s_{NN}}=200$ GeV.
We find a dominant influence of dynamical evolution over the initial conditions on the final observables.
\end{abstract}


\maketitle
\section{Introduction}

Ultrarelativistic heavy-ion collisions (HICs) at RHIC and LHC create a hot, ultra-dense state of deconfined quarks and gluons, known as the Quark-Gluon Plasma (QGP) \cite{Hirano:2005wx,Shuryak:2008eq, vanHees:2005wb, Song:2010mg, Asakawa:2006tc}. Experimental and theoretical studies seek to characterize the QGP, understand its formation, and explore its evolution and observable signatures.
HICs involve multiple stages, typically divided into an initial pre-equilibrium phase (for $t < 1$ fm/c) and a subsequent evolution phase covering expansion, thermalization, and hadronization. Various model approaches introduce different assumptions that have an impact on experimental interpretations.

Despite significant progress uncertainties remain, particularly in understanding thermalization and the transition from an out-of-equilibrium state to a thermalized QGP. The first phase involves the energy deposition and primary scattering, setting the initial conditions. The second phase includes the non-equilibrium evolution, leading to QGP formation and hydrodynamic expansion. The final stage covers hadronization, followed by a hadronic cascade and freeze-out.

Exploring different theoretical models and simulations for these stages is crucial. Several formalisms, including the color glass condensate (CGC) frameworks, classical Yang-Mills dynamics, and transport models, describe the initial stage \cite{Berges:2013eia,Kurkela:2018wud,Giacalone:2019ldn, Greif:2019ygb,Xu:2017pna}. For the QGP evolution relativistic viscous hydrodynamics effectively models collective flow, while transport and hybrid models bridge partonic and hadronic descriptions. Systematic comparisons of these approaches, validated against experimental data, are a key to advance our understanding of the QGP and the dynamical evolution of the strongly interacting matter. 

The goal of our study is to disentangle the impact of the Initial-State and the Evolution Dynamics on Final Observables in heavy-ion Collisions.
For that we use  EPOS4 (Energy conserving multiple Parallel scattering approach, accomodating factOrization and Saturation) \cite{werner:2023-epos4-overview, werner:2023-epos4-heavy, werner:2023-epos4-smatrix,werner:2023-epos4-micro, Werner:2024-pedagog} 
(replacing earlier EPOS versions \cite{Drescher:2000ha,Werner:2007bf,Werner:2010aa,Werner:2013tya}) 
and PHSD (Parton-Hadron-String Dynamics) microscopic approach \cite{Cassing:2008sv, Cassing:2009vt, Cassing:2008nn, Bratkovskaya:2011wp, Konchakovski:2011qa, Song:2015sfa, Bratkovskaya:2017gxq, Linnyk:2015rco, Moreau:2019vhw}; 
both models have well-defined "initial"  and  "evolution" stages: 
\begin{itemize}
\item
The initial phase of EPOS (EPOSi) amounts to an S-matrix approach for instantaneous parallel scatterings, whereas the following dynamics has been realized so far by assuming that a fast equilibration occurs followed by a hydrodynamical evolution (EPOSe).
\item
The PHSD models the initial high energy nucleon-nucleon scatterings and the string formation based on the LUND string model using the Pythia event generator in the initial phase (PHSDi), whereas the evolution is based on a microscopic covariant dynamical approach for strongly interacting systems using Kadanoff-Baym equations (PHSDe) in the semi-classical limit.
\end{itemize}

When comparing two models, such as EPOS and PHSD, it is not always straightforward to determine the relative contributions of the initial phase (i) and the expansion phase (e) to the final results. To address this issue, the idea of combining the initial phase from EPOS (EPOSi) with the evolution from PHSD (PHSDe) gave rise to the \textbf{EPOSir+PHSDe} model, where EPOSir adds a rope procedure to EPOSi. This approach allows for a clear comparison between two models based on distinct contributions from the initial conditions and the evolution.
By comparing EPOSir+PHSDe to  EPOS, we isolate the differences arising from the evolution phase, since both models share identical initial conditions but differ in their time evolution. On the other hand, when comparing EPOSir+PHSDe to  PHSD, we are comparing models with different initial conditions but the same evolution, allowing us to clearly distinguish between the contributions of the initial conditions and the evolution phase. 

In this study we focus on transverse momentum spectra and flow harmonics in Au+Au collisions at the invariant energy $\sqrt{s_{NN}}=200$ GeV, which is well suited to study the dynamics of strongly interacting matter due to the large volume of the QGP and relatively long life time of the formed fireball such that the system proceeds towards a partial thermalization.

The outline of the paper is as follows: in Section II we present the theoretical basis of EPOS and PHSD and the combined  new approach, EPOSir+PHSDe. In Section III we present and compare the results of the three alternative approaches, EPOS, EPOSir+PHSDe, and PHSD, for various observables of Au+Au collisions at the top RHIC energy, such as bulk matter observables, and anisotropic flows. We summarize our findings in Section IV.


\section{Combined EPOS+PHSD approach}

\subsection{EPOS}

EPOS (which refers to EPOS4 in this paper) starts with so-called primary scatterings, happening simultaneously at $t=0$  (EPOSi).  It follows  a core-corona procedure, an assumed fast equilibration of the core with a subsequent hydrodynamical evolution and microcanonical  hadronization, and finally  a hadronic cascade (EPOSe). For a detailed description, see Refs. \cite{werner:2023-epos4-overview, werner:2023-epos4-heavy, werner:2023-epos4-smatrix,werner:2023-epos4-micro}; a pedagogical overview can be found in Ref. \cite{Werner:2024-pedagog}.

\begin{figure}[h!]
    \centering
    \includegraphics[scale=0.56]{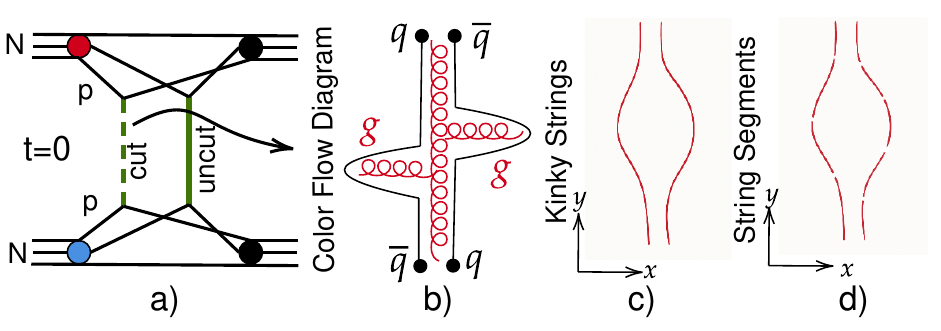}
    \caption{From Pomerons to string segments: a) Elementary interactions for a nucleon-nucleon scattering in EPOS, showing a cut Pomeron (dashed green line) and an uncut Pomeron (full green line). b) Color flow diagram corresponding to a cut ladder. c) Kinky strings from color flow diagram. (d) String segments.}
    \label{pomeron-stringsegments1}
\end{figure}

\textbf{EPOSi:} 
The primary scatterings amount to parallel partonic scatterings, happening instantaneously, at $t=0$.
The theoretical tool is S-matrix theory, using a particular form of the multiple scattering S-matrix. 
Each elementary scattering process is described by Pomeron exchange, a Pomeron being realized as a parton ladder \cite{werner:2023-epos4-heavy}.
A very important aspect of EPOS is the treatment of high parton density effects, representing nonlinear parton evolutions, which are accounted for by implementing a dynamical saturation scale, which depends for each Pomeron on its energy, but also the environment (neighboring Pomeons) \cite{werner:2023-epos4-smatrix}. The approach is compatible with the current understanding of parton saturation.
To evaluate the multiple scattering S-matrix expressions, cutting rule techniques are employed \cite{werner:2023-epos4-smatrix}, which leads to an expansion of cross sections in terms of products of cut and uncut Pomerons (sketched in Fig. \ref{pomeron-stringsegments1}a). 
A cut Pomeron is considered to be an essentially longitudinal color field in the form of a flux tube, whose dynamics is described by relativistic string theory via so-called kinky strings, see Figs. \ref{pomeron-stringsegments1}b and c.
In elementary collisions, the string breaking by quark-antiquark production leads to hadron formation from the individual string segments identified  as hadrons, see Fig. \ref{pomeron-stringsegments1}d.

To construct "the core", one considers the space-time evolution of these string segments and in particular the intersection of their trajectories with a hyperbola representing a fixed proper time  $\tau$. We know not only the space-time position of each segment but also its velocity. In principle, we have a dynamical system, where all elements move, but we may get some estimate of the fate of each segment by considering its trajectory through the matter composed of all the other segments at their fixed position at $\tau$. In this way, we compute the energy loss of each string segment as \cite{werner:2023-epos4-micro}:
\begin{equation}
P_{t}^{new} = P_{t} - f_{Eloss} \int_{\gamma} \rho dL,
\end{equation}
where $\gamma$ and $P_{t}$ are the trajectory and transverse momentum of the string segment, respectively. $f_{Eloss}$ is a nonzero constant, and $\rho$ is the local string density. 
If  $P_{t}^{new}$ is positive, the string segment can escape from the dense area, and it is categorized as a \textbf{"corona particle"}. If $P_{t}^{new}$ is negative, however, the string segment lost all its energy and is unable to leave, thus it will remain in the dense area and is called \textbf{"core particle"}. All the core particles constitute "the core".

\textbf{EPOSe:} Here we have to distinguish between the core and the corona particles. 
The core expands hydrodynamically based on relativistic viscous hydrodynamic equations with shear viscosity to entropy ratio $\eta /s = 0.08$ and zero bulk viscosity \cite{Werner:2013tya,werner:2023-epos4-micro}. 
The microcanonical method \cite{werner:2023-epos4-micro} is employed to  hadronize   the core. The corresponding hadrons, as well as the corona particles (originating directly from string decay)   follow straight-line trajectories, but they continue to interact via hadronic scattering, employing the UrQMD model \cite{Bleicher:1999xi}.

\begin{figure}[h!]
    \centering
    \includegraphics[scale=0.36]{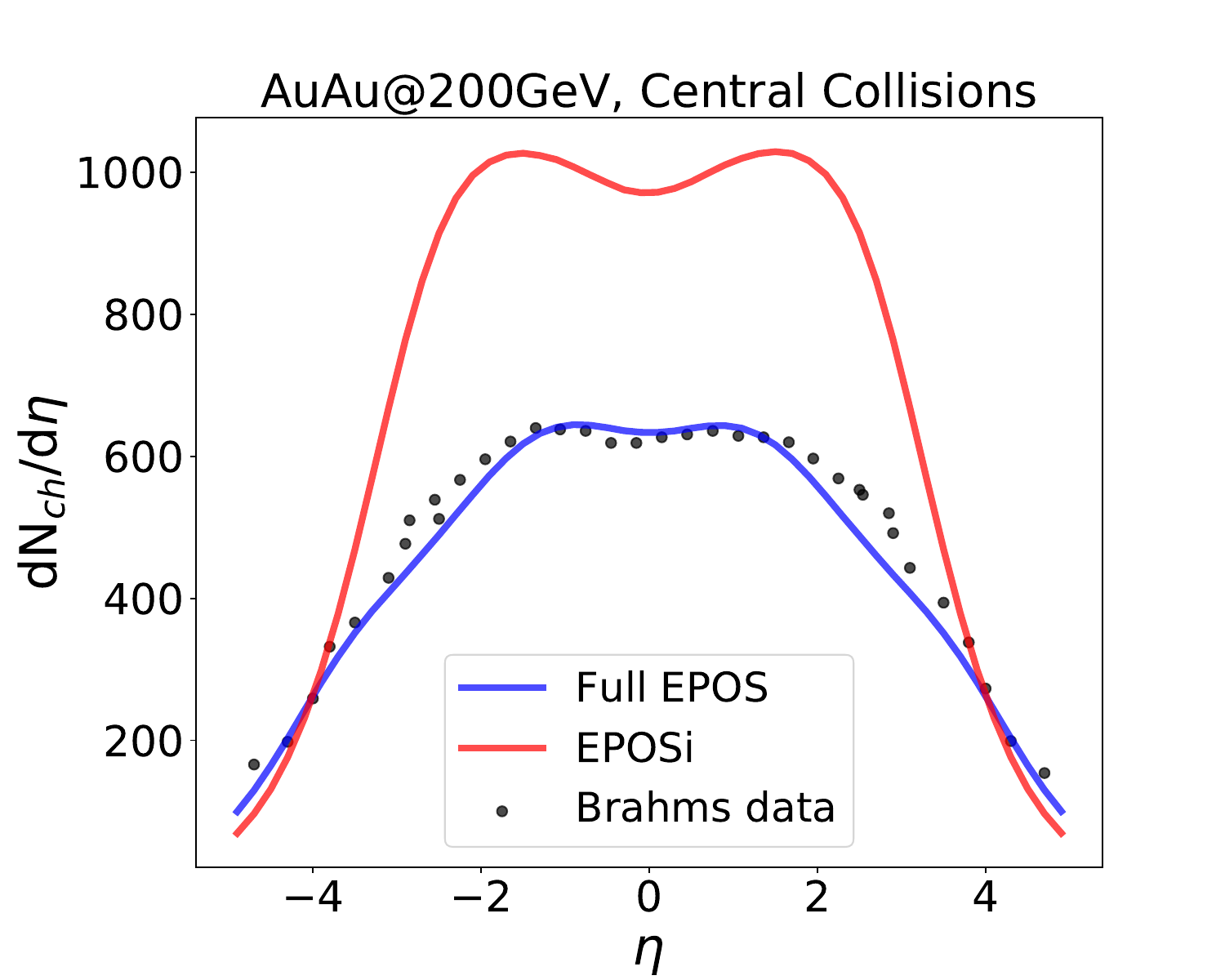}
    \caption{Charged particle multiplicities ($dN_{ch}/d\eta$) as a function of pseudorapidity ($\eta$) in Au+Au collisions at invariant energy 200 GeV for different simulations, Full EPOS (blue curve, including hydro), and EPOSi (red curve, without hydro, without hadronic cascade) for central collisions (0-5$\%$). The experimental data are taken from BRAHMS \cite{BRAHMS:2001llo} with black points.} 
    \label{eta-epos}
\end{figure}

A full EPOS simulation amounts to both, primary and secondary interactions (EPOSi and EPOSe).
But it is interesting to see what we get (concerning particle production) if we consider only EPOSi, assuming that all string segments convert to hadrons.
The charged particle multiplicity  is shown in Fig. \ref{eta-epos}  as a  function of the pseudorapidity for full EPOS (including hydro) and just EPOSi (without hydro),  
compared to data from the BRAHMS experiment \cite{BRAHMS:2001llo}. 
Full EPOS (with hydro, blue curve) can reasonably well reproduce  the experimental data. However, EPOSi (without hydro, red curve) overestimates the experimental data  by a large amount. 
This shows the essential role of hydrodynamics, which allows to redistribute the system energy from the "mass production" to the kinetic energy of the expanding fireball.

\subsection{PHSD}

The Parton-Hadron-String Dynamics (PHSD) is a non-equilibrium microscopic transport approach that incorporates both hadronic and partonic degrees of freedom \cite{Cassing:2008sv, Cassing:2008nn, Cassing:2009vt, Bratkovskaya:2011wp, Konchakovski:2011qa, Linnyk:2015rco, Moreau:2019vhw}. It provides a comprehensive description of the entire evolution of relativistic heavy-ion collisions, starting from initial out-of-equilibrium nucleon-nucleon ($NN$) interactions, progressing through the formation and interactions of the quark-gluon plasma, and extending to hadronization and the final-state interactions of produced hadrons.  

The dynamical evolution of the system is governed by the Cassing-Juchem generalized off-shell transport equations, formulated within the test-particle representation \cite{Cassing:1999wx, Cassing:1999mh}. These equations arise from a first-order gradient expansion of the Kadanoff-Baym equations \cite{KadanoffBaym}, which effectively describe strongly interacting degrees of freedom \cite{Juchem:2004cs, Cassing:2008nn}. For a more detailed discussion, see \cite{Cassing:2021fkc}.

\textbf{PHSDi:} The initial collisions of PHSD are build by the primary independent scattering of the initial nucleons (according to Wood-Saxon distribution) from colliding nuclei. The high energy $NN$ reactions (as well as elementary baryon-baryon ($BB$) reactions above $\sqrt{s_{BB}^{th}} = 2.65$ GeV, meson-baryon ($mB$) collisions above $\sqrt{s_{mB}^{th}} = 2.35$ GeV, and meson-meson ($mm$) reactions above $\sqrt{s_{mm}^{th}} = 1.3$ GeV) leading to multi-particle production  are described by the Lund string model \cite{NilssonAlmqvist:1986rx}. These are realized through event generators FRITIOF 7.02 \cite{NilssonAlmqvist:1986rx, Andersson:1992iq} and PYTHIA 6.4 \cite{Sjostrand:2006za}, which are "tuned" for a more accurate description of low and intermediate energy reactions. 
The "PHSD tune" \cite{Kireyeu:2020wou} of the Lund string model modifies string fragmentation and hadron properties in the hot, dense medium of heavy-ion collisions. It incorporates chiral symmetry restoration via the Schwinger mechanism \cite{Cassing:2015owa, Palmese:2016rtq}, accounts for the initial-state Cronin effect in $k_T$ broadening, and implements in-medium hadron properties by using momentum-, density-, and temperature-dependent spectral functions instead of constant-width non-relativistic ones \cite{Bratkovskaya:2007jk, Song:2020clw}.
The color neutral strings decay into leading hadrons and "prehadrons," unformed mesons and baryons from string fragmentation in the Lund model. 

\textbf{PHSDe:} 
The prehadron formation time is \( t_F = \tau_F \gamma \), with \( \tau_F = 0.8 \) fm/\( c \) and \( \gamma = 1/\sqrt{1 - v^2/c^2} \). Leading hadrons interact immediately with reduced cross sections \cite{Cassing:1999es}, while prehadrons remain non-interacting for \( t < t_F \) in "cold" matter with energy density \( \varepsilon < \varepsilon_C \), where \( \varepsilon_C = 0.5 \) GeV/fm\(^3\), based on lattice QCD \cite{Borsanyi:2022qlh}. Below \( \varepsilon_C \), prehadrons evolve into hadronic states and interact via hadronic cross-sections. 

If \( \varepsilon \simeq \varepsilon_C \), prehadrons melt into thermal partons, governed by the Dynamical Quasi-Particle Model (DQPM) model.
Partonic interactions in the QGP phase are modeled within the effective DQPM model, formulated in the two-particle irreducible (2PI) propagator representation \cite{Cassing:2007yg, Cassing:2007nb, Moreau:2019vhw, Soloveva:2019xph}. The DQPM describes non-perturbative QCD matter in line with the lattice QCD equation-of-state (EoS) at finite temperature \( T \) and baryon chemical potential \( \mu_{\mathrm{B}} \) \cite{Aoki:2009sc, Cheng:2007jq}.  
In this approach, partons are off-shell quasiparticles with complex self-energies: the real part determines a dynamically generated mass, while the imaginary part governs interaction rates, yielding spectral functions with finite widths. The \( (T, \mu_{\mathrm{B}}) \)-dependence of self-energies follows a hard-thermal-loop-inspired parametrization, with a single parameter fixed by matching the DQPM entropy density (or pressure/energy density) to lattice QCD results at \( \mu_{\mathrm{B}}=0 \). The partonic interactions include elastic \(2 \leftrightarrow 2\) scatterings (\(qq \leftrightarrow qq\), \(\bar{q} \bar{q} \leftrightarrow \bar{q} \bar{q}\), \(gg \leftrightarrow gg\)) and inelastic \(2 \leftrightarrow 1\) processes (\(gg \leftrightarrow g\), \(q\bar{q} \leftrightarrow g\)), described by a relativistic Breit–Wigner cross section.
The DQPM framework provides a consistent description of the QCD thermodynamics and the QGP properties in terms of transport coefficients in the whole $(T,\mu_B)$ plane \cite{Moreau:2019vhw, Soloveva:2019xph, Fotakis:2021diq, Soloveva:2020ozg, Grishmanovskii:2023gog} as well as a non-equilibrium description - via Kadanoff-Baym equations - of strongly interacting degrees of freedom propagated in a self-generated scalar mean-field potential \cite{Cassing:2009vt}.
We note that in this study we use the PHSD version 4.1 which is linked to the DQPM model for $\mu_B=0$, which is a good approximation for the relativistic energies considered here.

As the fireball expands, parton hadronization occurs near the phase boundary between hadronic and partonic matter, which lattice QCD data indicate as a crossover \cite{Borsanyi:2022qlh}. The hadronic system is then described using off-shell HSD dynamics, optionally incorporating self-energies for hadronic degrees of freedom \cite{Cassing:2003vz, Bratkovskaya:2007jk, Song:2020clw}.

\subsection{Reasoning for combining EPOS+PHSD approaches}

As described above, the EPOS and PHSD approaches have fundamentally different initial conditions and matter evolutions, nevertheless, they provide rather similar results in many observables, but differences in  some others. 

In order to disentangle the influence of the initial state and evolution effects on the final observables of heavy-ion collisions - which is the aim of this study - we combine the two different approaches in a single framework, called "EPOSir+PHSDe", in which the standard EPOS initial conditions (EPOSi), complemented by a rope procedure (EPOSir), are employed as the starting point for the  PHSD-based parton and hadron evolution (PHSDe).

\section{Dynamics of HICs within EPOSir+PHSDe}

\subsection{Initial conditions according to EPOSi}

The combined EPOSir+PHSDe approach is based on EPOS initial conditions discussed in Section II.A.
thus, here we focus only on the issues which are important for the EPOS/PHSD connection.

\begin{figure}
    \centering
    \includegraphics[scale=0.37]{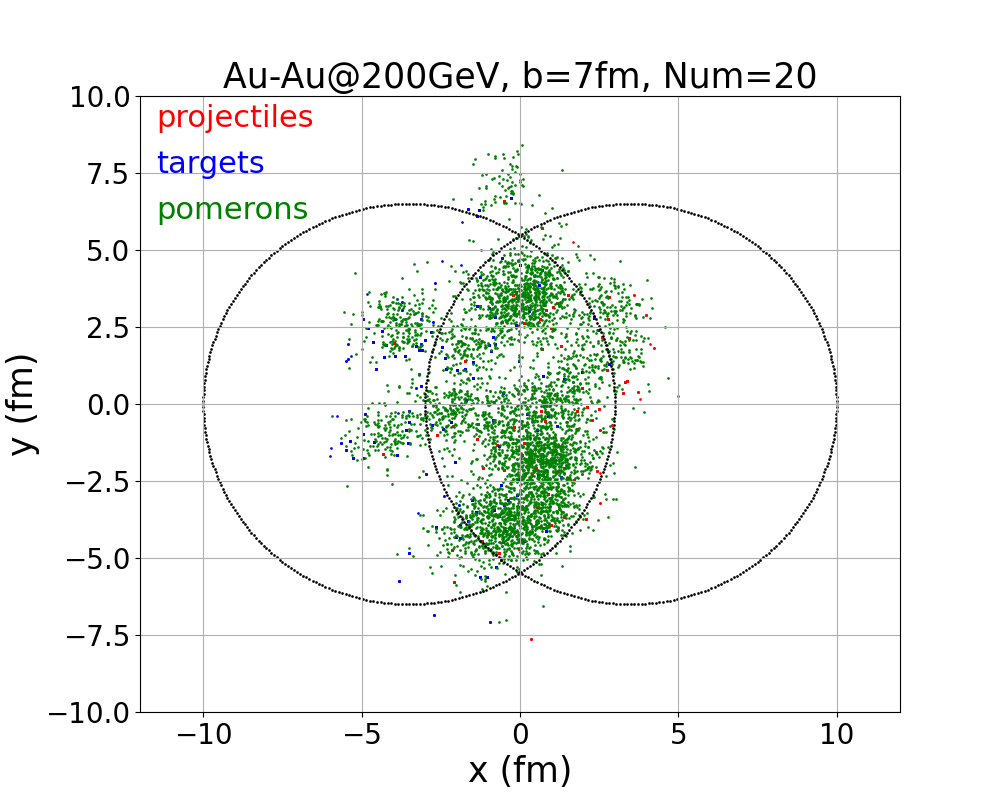}
    \caption{The positions in the transverse plane of the projectile (red points) and target (blue points) participants and their corresponding Pomerons. We consider here semi-peripheral Au+Au collisions at the invariant energy $\sqrt{s_{NN}}=200$ GeV.  }
    \label{av}
\end{figure}

\begin{figure}
    \centering
    \includegraphics[scale=0.65]{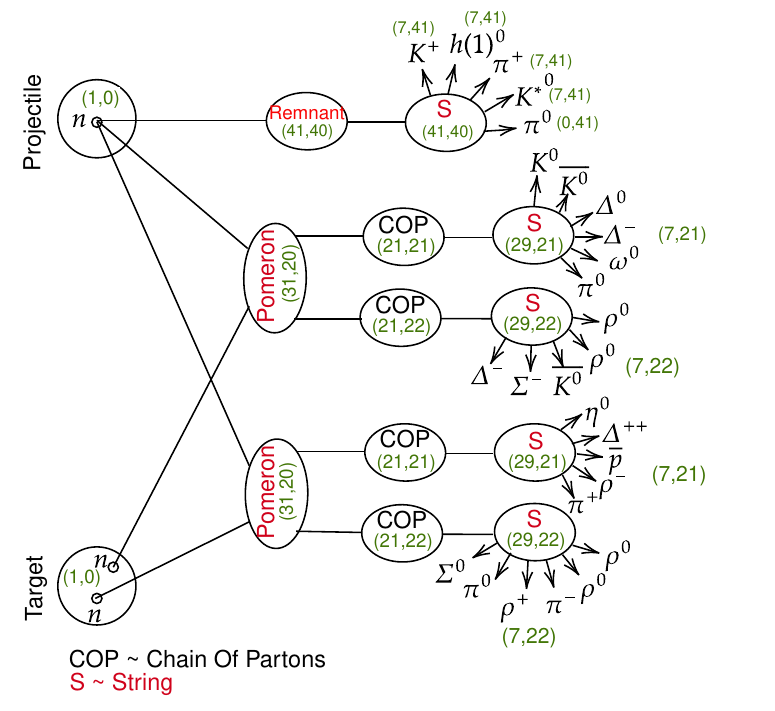}
    \caption{Sketch of initialization in EPOSi based on the multiple Pomeron exchange. Pomerons represent the interactions between nucleons from the projectile and the target. The Pomerons convert into several chains of partons (COP), and then the COP is split into string segments.
    The first and second numbers in parentheses represent the particle's status (ist) and type (ity).}
    \label{pomeron-exchange}
\end{figure}

As the first step, we use the known distribution of nucleons in the initial nuclei (Wood-Saxon in case of  heavy nuclei such as Au) to generate nucleon positions for both projectile and target. 
In EPOSi, an S-matrix approach allows for the determination of pairs of interacting nucleons and the corresponding Pomerons.  Interacting nucleons are referred to as participants, the other ones are called spectators. The participants are mainly within the overlap zone, which is a theoretical construction representing the overlap of two spheres, see Fig. \ref{av}.  The Pomerons are identified by parton ladders. Following the color flow, one constructs chains of partons (COP), which can be mapped to kinky strings. Two integer numbers will characterize the different "objects", referred to as "status" (ist) and "type" (ity). The first one allows differentiating between Pomerons, parton chains, and string segments, whereas the latter one gives additional information, like being core or corona particle.  
The whole picture, namely interactions via Pomerons, identification of chains of partons, kinky strings, and string segments, is sketched in  Fig.  \ref{pomeron-exchange}. 

\begin{figure}
    \centering
    \includegraphics[scale=0.45]{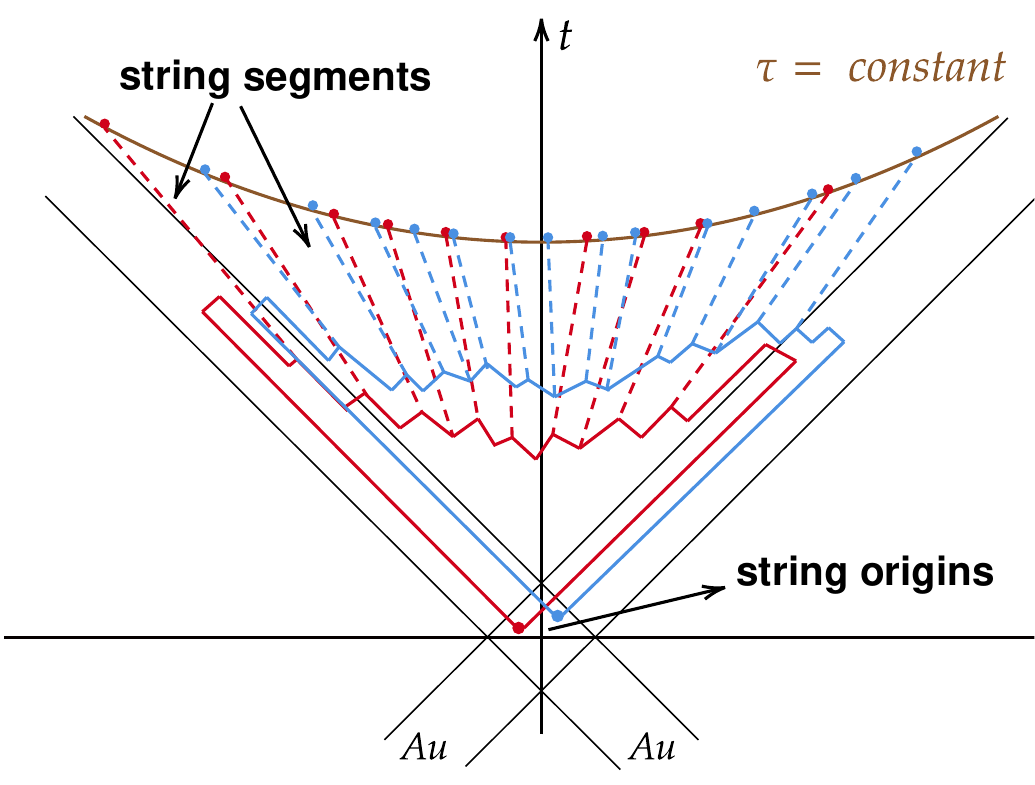}
    \caption{Space-time picture of the string segment production (see text).}
    \label{lightcone2}
\end{figure}

Technically, the method consists of three steps: first, the Pomerons produce several COPs, then they are converted to kinky strings (S), and finally, the strings are split into several string segments, as discussed in the EPOSi part. In Fig. \ref{lightcone2}, we show the corresponding space-time picture. We show for simplicity a situation where we have just two strings (one in blue, the other one in red). The red and blue points close to the $x$-axis represent the locations of the string origins. Starting at these points, the strings evolve (here we sketch simple yo-yo strings) and then break. The trajectories of the string segments (also referred to as prehadrons) are indicated as dashed lines (red and blue from the two different strings). We indicate as dots the intersections of the string segment trajectories with a hyperbola corresponding to some proper time $\tau$.

At the same time, we have projectile and target remnants that decay  into several "prehadrons". For simplicity, we show only the projectile remnant in Fig. \ref{pomeron-exchange}. 

\begin{figure}
	\begin{minipage}{0.45\textwidth}
		\centering
        \includegraphics[scale=0.6]{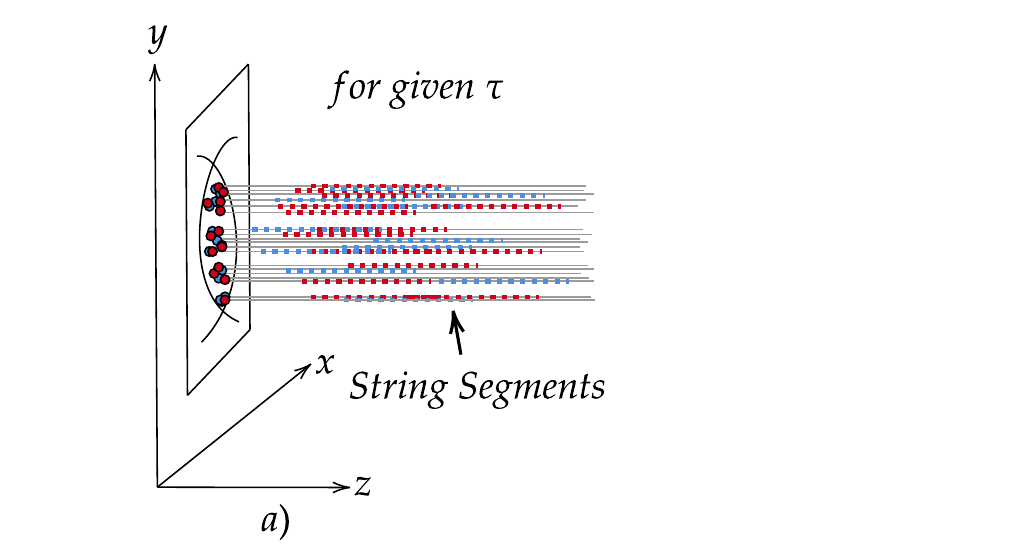}
	\end{minipage}\\
	\begin{minipage}{0.45\textwidth}
		\centering
        \includegraphics[scale=0.6]{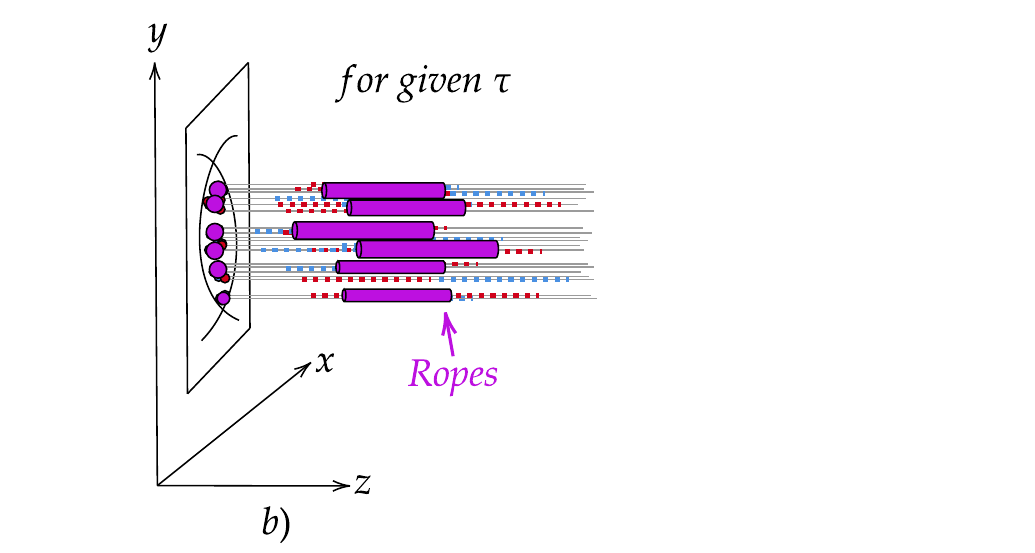}
	\end{minipage}\\
	\begin{minipage}{0.45\textwidth}
		\centering
		\includegraphics[scale=0.68]{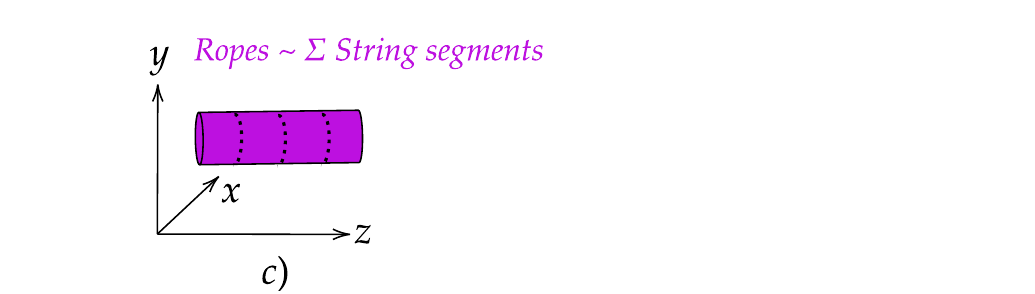}
	\end{minipage}\\
	\begin{minipage}{0.45\textwidth}
		\centering
		\includegraphics[scale=0.68]{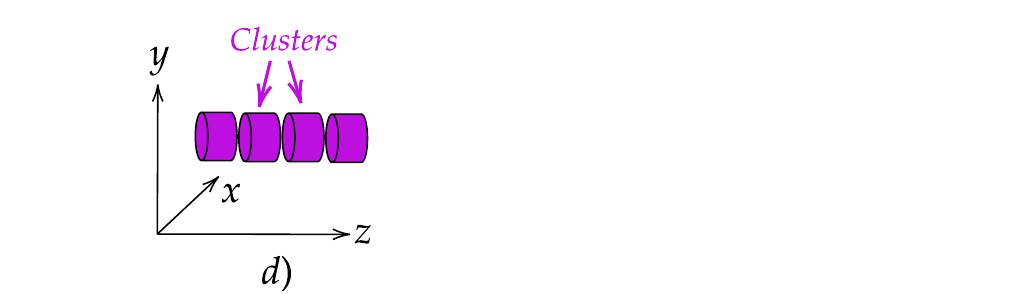}
	\end{minipage}
    \caption{
    From string segments (prehadrons)  to cluster formation in three dimensions at given $\tau$. a) String segments production. b) Overlapping string segments defining the rope segments. c) Constructing the rope segment. d) Cutting the rope segment into disks (clusters).  }
	\label{rope11}
\end{figure} 

The prehadrons may be very close to each other, this is why in EPOS a core-corona procedure is employed. In EPOS, the core segments define the initial condition for a hydrodynamical evolution. The string segments are shown in three dimensions, i.e. transverse plane versus longitudinal direction $z$ in Fig. \ref{rope11}a, employing the space-time rapidity  $\eta_{s}$ instead of $z$ for the longitudinal dimension,  being defined as
$\eta_{s}= \frac{1}{2} \log ((t+z)(t-z)).$

Fig. \ref{rope11}a depicts the string segments at a given $\tau$. The core segments are presented as colored (blue and red) objects. For simplicity, only flat strings are shown. It is also possible (but not shown) to have corona particles in the middle of the core segments,   if their kinetic energy is large enough.

\subsection{EPOSir:  EPOSi + ropes}

In EPOS+PHSD, we also use this core-corona picture. Having identified the core, we have to transform it into an initial condition of PHSD, which requires "prehadrons".  Naively, one may directly use all the "EPOS prehadrons" as prehadrons for PHSD. But this turned out not to work at all, the multiplicity will be much too high. So we  modify EPOSi by introducing ropes, referred to as "EPOSi+ropes" or simply EPOS4ir. We consider connected areas of the core as  "ropes", which generally defines objects obtained from fusing strings. This is essentially what we do with the core construction. All the other string segments (those not contributing to the ropes) are referred to as corona particles. We sketch the "rope production" in Fig. \ref{rope11}b. 

\begin{figure}[tp]
    \centering
    \includegraphics[scale=0.36]{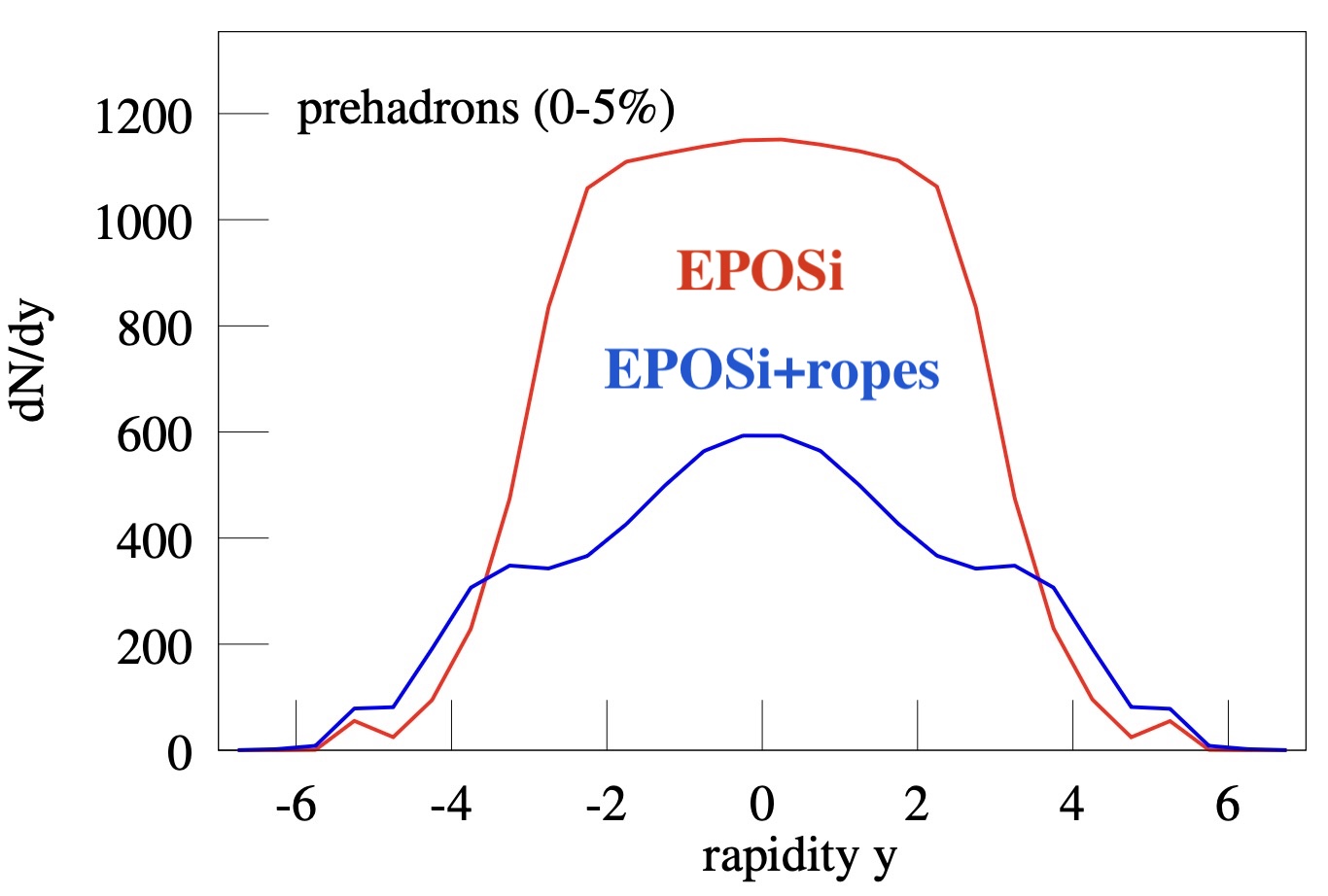}
    \caption{ Initial stage rapidity distributions of prehadrons for EPOSi (red curve) and EPOSi+ropes (=EPOSir, blue curve) in most central 0-5\% Au+Au collisions at $\sqrt{s_{NN}}=200$ GeV averaged over 5000 events.}
    \label{eposi+ropes}
\end{figure}

We point out that using  EPOSi leads to the production of many prehadrons and, consequently, to a strong overestimation of the final hadron yield (roughly twice as many as in experiments), when employing PHSDe. 
As discussed earlier, this “overproduction” is compensated by the  hydro evolution, which helps to convert the energy stored in the masses into flow and work.
The  introduction of the ropes, i.e. the method "EPOSi+ropes" (EPOSir in short),  decreases the number of prehadrons (per unit of pseudo-rapidity), as shown in Fig \ref{eposi+ropes}  for central Au+Au collisions at $\sqrt{s_{NN}}=200$ GeV. 
Thus, the ropes allow to convert the initial large mass production in the Pomeron picture in EPOS to the kinetic energy of degrees-of-freedom.

Ropes are considered as longitudinal color fields similar to ordinary strings. 
We then break the rope segments into several pieces, called "clusters", for technical reasons, as illustrated in Fig. \ref{rope11}c.

Technically, we employ a cluster algorithm, based on a cell-centered grid in three dimensions, which enables us to define slices in the longitudinal variable and identify connected transverse areas and the corresponding transverse density distribution. These connected sections in transverse space amount to rope slices, or clusters, as illustrated in Fig. \ref{rope11}d.

These rope slices (or clusters) are not static, but they are assumed to have a transverse flow profile  (resulting in higher transverse momenta of its decay products) and  have a longitudinal flow profile (resulting in a broader rapidity distribution).  Both effects will reduce the particle multiplicity of particles from cluster decay.

To prepare the evolution for PHSD, we will decay the clusters into hadrons, using the usual EPOS procedure of microcanonical decay of an object of mass $M$ (with flow profiles) in its center of mass, with a subsequent boost back into the lab system.  The technical details can be found in \cite{werner:2023-epos4-micro}.
In EPOSir+PHSDe language, we call these hadrons \textbf{"rope core prehadrons"}  (they have hadron properties, but are not final state hadrons yet).

The pre-hadronization is realized slice by slice, where different slices correspond to different values of $\eta_{s}$.  There is a strong correlation between $\eta_s$ and the rapidity $y$ of the rope segment, in the sense that the rapidity of the cluster is essentially equal to $\eta_s$. Therefore, the core prehadrons produced from the decay of a rope slice at given $\eta_s$ show up at $y$ close to $\eta_s$.

The sequence of clusters corresponds to a sequence of $\eta_s$ values (in an ordered fashion) and correspondingly the clusters are sitting at different values of rapidity (again ordered), so the final rapidity distribution of prehadrons will be a sum of many  (Gaussian-like) peaks, adding up to a broad (essentially flat) distribution.  

To avoid possible confusion, let us recall the definitions of the different core prehadrons: those produced in the EPOS core-corona procedure are called "EPOS core prehadrons", which are the basis of the production of ropes and clusters. The latter decay, and the decay products are called "rope core prehadrons", because they are produced from rope decay. So at this stage, we have "corona prehadrons" and "rope core prehadrons".

 \subsection{Inserting prehadrons into PHSD}

\begin{figure}[h!]
    \centering
    \includegraphics[scale=0.45]{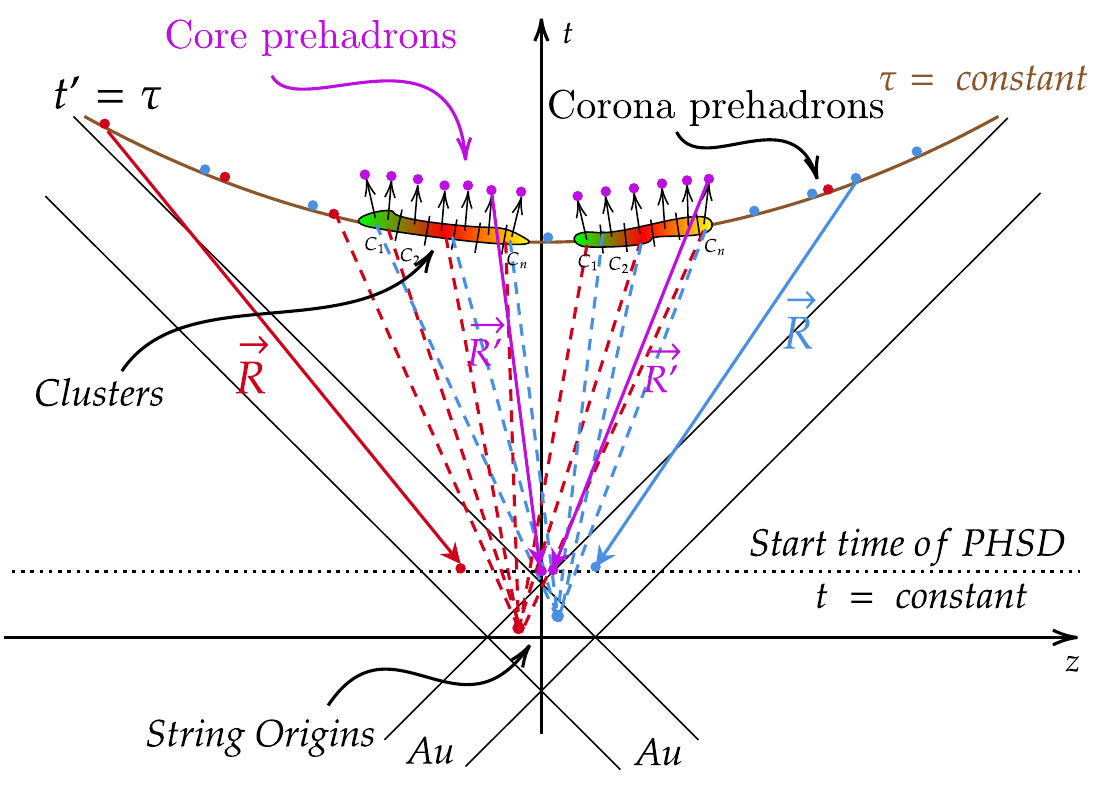}
    \caption{Extrapolation back-in-time procedure of core and corona prehadrons to the starting time of PHSD.}
    \label{extrapolation-phsd}
\end{figure}

 On one hand, EPOS operates with Milne coordinates, and thereby the prehadrons are produced on a hyperbola in space-time. On the other hand, the PHSD code uses the cartesian coordinates $(x,y,z)$ at a given time $t$. The knowledge of the space distribution of energy and baryon densities is crucial for PHSD dynamics. Therefore one has to appoint some relation between the two coordinate systems.
 In principle, we just need to extrapolate the prehadrons from the hyperbola back to the constant timeline, see Fig. \ref{extrapolation-phsd}. Since the positions, time, and momenta are known, then we know the trajectories:
\begin{equation} \label{eqn:extrapolation}
    \vec{R}(t)=\vec{R}(t_p)+\vec{V_p} \times (t-t_{p}),
\end{equation}
where $\vec{R}(t_p)$ in the position of the particle at time $t_p$, $V_p$ -- its velocity. Thereby, one can compute the position at the initial time in PHSD ($t=t_{ini_{PHSD}}$).  This is actually what is done for "corona prehadrons". 
They can be immediately inserted into PHSD.
We call this procedure "normal extrapolation", which can be seen by red and blue arrows in Fig. \ref{extrapolation-phsd}. 
We note that in PHSD the inserted prehadrons become formed hadrons after passing the formation time $t_F$ as discussed in Section II.B.

Concerning the "rope core prehadrons", the situation is more complicated, since these "prehadrons" did not exist prior to the hyperbola line, but what exists are their "parents".
The cluster decay particles (rope core prehadrons) are placed at the corresponding time at the positions of string segments having produced the clusters. These segments are referred to as "parents" with respect to the clusters, whereas the decay products (rope core prehadrons) are referred to as "children". All the string segments contributing to cluster formation are "EPOS core prehadrons". This nontrivial extrapolation procedure, namely placing the children (with their momenta) at the position of the parents, see the violet arrows in Fig \ref{extrapolation-phsd}, assures an accurate energy density distribution at any PHSD time step. The extrapolated children  are referred to as "PHSD core prehadrons".

The evolution of partonic and hadronic phases in PHSD is governed by the local energy density calculated on the space grid at each time step. The original algorithm was extended to include "parents" and "corona prehadrons".
If the average energy density exceeds the critical energy density (0.5 GeV/fm$^3$), the corresponding  PHSD core prehadrons must be melted before their formation into the partonic phase. This is called the "melting procedure".
Note that the corona prehadrons do not melt into the partonic phase but contribute to the energy density. 

\begin{figure}[h!]
    \centering
    \includegraphics[scale=0.45]{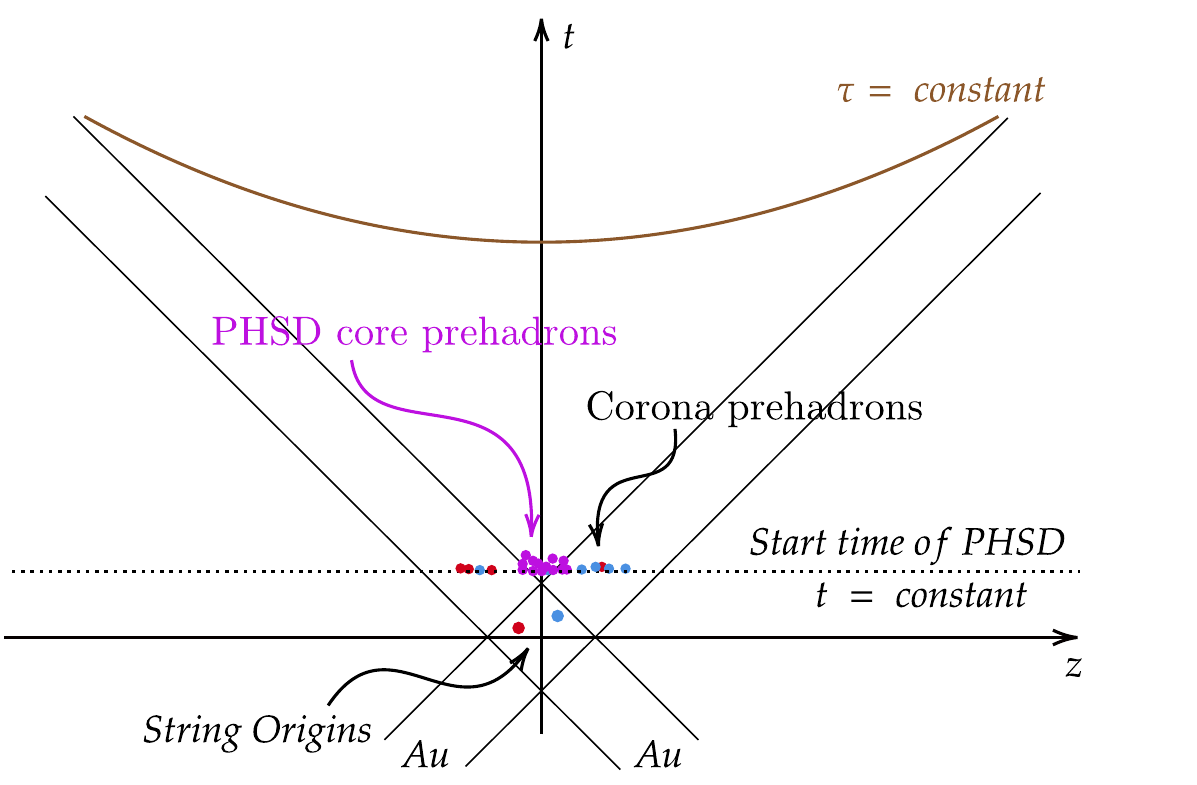}
    \caption{The schematic depiction of the final places of PHSD core prehadrons  and the corona prehadrons after extrapolation and melting at the start time of the PHSD evolution.  }
    \label{final-positions}
\end{figure}

Fig. \ref{final-positions} shows a schematic picture of the final positions of core and corona prehadrons following the extrapolation and melting processes to the PHSD start time. For the energy under consideration almost all of the prehadrons are melted during the first few PHSD time steps but for low energies, the process can be stretched in time.
The rest of the evolution of matter is carried out using the PHSD non-equilibrium dynamics once all prehadrons have been inserted into PHSD arrays.
In the following subsection, we will discuss the space-time evolution of particles in EPOSir+PHSDe.

\subsection{Space-time evolution of HICs in EPOSir+PHSDe}

PHSD employs a space-time grid that expands in time to describe HICs.
Since the particle density is high at the initial stage of the PHSD evolution, a considerable number of interactions occur. Therefore, the time step must be short enough to consider their interactions reasonably. The system expands later, and the size of the cells steadily rises.
In the axis parallel to the beam direction (generally referred to as the $z$-axis), the PHSD grids are expanded with time in order to cover the whole system on the grid.

Following the melting condition, the PHSD core prehadrons become deconfined into partons at a time of 0.064 fm/c.
The partonic phase begins at this time, and many partons exist in the overlap region. 
The partonic phase evolves in time until the system is cooling due to the expansion. Then the hadronization occurs dynamically for the quarks/antiquarks entering in the cells with a local energy density below the critical one. The quarks/antiquarks fuse into mesons or (anti)baryons according to the transition rate \cite{Cassing:2009vt}.
 In parallel, the number of mesons (baryons) rises with time due to hadronization, and they spread in space, resulting in blue (green) points.

\subsection{ Momentum eccentricity}\label{momentum_eccentricity}

\begin{figure}[h!]
    \centering
    \includegraphics[width=1.\linewidth]{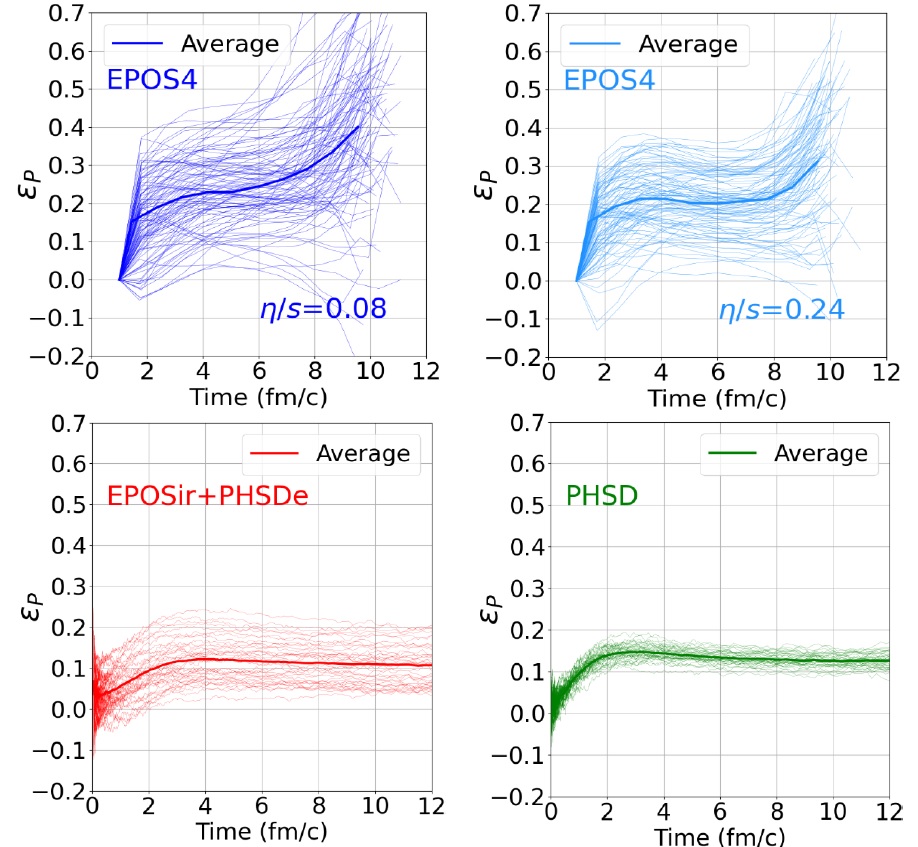}    
    \includegraphics[width=0.95\linewidth]{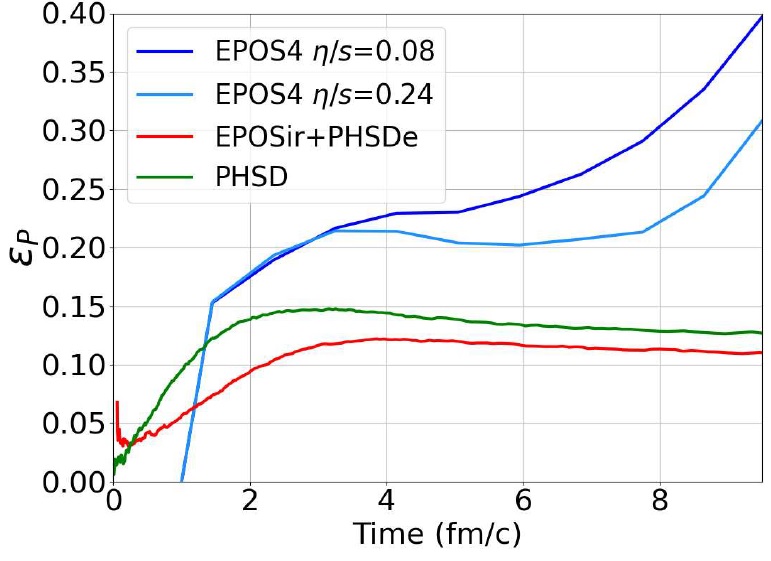}       
    \caption{The momentum eccentricity ($\epsilon_P$) as a function of time for EPOS with $\eta/s=0.08$ (top left), EPOS with $\eta/s=0.24$ (top right), PHSD (middle right), EPOSir+PHSDe (middle left) calculated for 100 EPOS events,  800 EPOSir+PHSDe and 800 PHSD events of Au+Au collisions at $\sqrt{s_{NN}}=200$ GeV with an impact parameter b=7 fm, as well as a comparison of the averaged values of $\epsilon_P$ (bottom). }     
    \label{eccp-compare}
\end{figure}

In order to understand better the dynamical observables of the bulk matter, we investigate here the time evolution of the momentum eccentricity for three models - EPOS, PHSD and EPOSir+PHSDe which shows how the pressure gradients develop in time.

The momentum eccentricity of each cell is defined as 
\begin{equation}
    \epsilon_p=\frac{\sum \varepsilon \ v^2 \cos(2\phi)}{\sum \varepsilon \ v^2},
    \label{eccx-formula}
\end{equation}
where $v^2=v_x^2+v_y^2$ is the transverse hydrodynamic velocity of the cell, $\phi=\arctan{(v_y/v_x)}$ the azimuthal angle, and  $\varepsilon$  the energy density of the cell.
The hydrodynamical velocity is defined as $u^{\mu} T_{\mu \nu} = \varepsilon u_{\nu}$, where $T_{\mu \nu}$ is the energy-momentum tensor; the eigenvector $u^\mu$ is the four-velocity ($v=u/u^0$). 
To compute the energy density, we use the energy-momentum tensor $T^{\mu \nu}$ from kinetic theory, which is given by \cite{Drescher:2000ec}:
\begin{equation}
    T^{\mu \nu}(\vec{q}) = \int \frac{d^{3}p}{E} p^{\mu} p^{\nu} f(\vec{q},\vec{p}),
\end{equation}
where $\vec{q}$ is a position vector, $\vec{p}$ indicates a momentum vector, and $f$ denotes the phase space density for a given time.
The energy density is given as $T^{00}$ in the comoving frame. This is the way one determines the energy density in EPOS, more precisely by solving $u^{\mu} T_{\mu \nu} = \varepsilon u_{\nu}$.

In both EPOSir+PHSDe and PHSD, we  compute the energy density  as \cite{Marty:2014zka}:
\begin{equation}
    \varepsilon= \frac{\sum_{i} E_i}{V_{cell}},
\end{equation}
where the sum of all particles in each cell is provided by $\sum_{i}$, and the volume of the cell is defined by $V_{cell}=\Delta x \Delta y \Delta z$. $\Delta x=\Delta y=$1 fm which is equal to the hadron size and $\Delta z= 1/\gamma_{com}$ fm. $\gamma_{com}$ is the Lorentz $\gamma$ factor for the transformation into the center-of-mass of the colliding nuclei.
Therefore, the energy density of the cells is determined by 
\begin{equation}
    \varepsilon'=\frac{E'}{V'}=\frac{E/\gamma}{V \times \gamma}=\frac{\varepsilon}{\gamma^{2}}.
    \label{EDrestframe}
\end{equation}

In Fig. \ref{eccp-compare} we show the momentum eccentricity $\epsilon_P$ for semiperipheral Au+Au collisions at the invariant energy $\sqrt{s_{NN}}=200$ GeV for the three models.  The momentum eccentricity in EPOS using hydro with the default value (see Sec. II.A and  \cite{Werner:2013tya,werner:2023-epos4-micro}) of  the shear viscosity to entropy ratio $\eta/s =0.08$  (top left panel)  strongly increases on average (thick line) with time, but with large fluctuations for individual events (thin lines): some of these lines grow very strongly, beyond values of 0.7, and others show a modest increase. 
This spread is a consequence of the large fluctuations in the initial shape: sometimes an elongated shape is produced, and sometimes it is completely symmetric.
The former leads to a strong asymmetric flow due to hydrodynamic pressure and a strong increase in the momentum eccentricity. The latter leads to a small asymmetric flow and a modest increase in momentum eccentricity.
Since elongated shapes are produced frequently, they affect the average curve considerably, and one gets the observed strong increase of the average momentum eccentricity with time.  

We show as well the EPOS result for  $\eta/s =0.24$  (the top right panel), where the average curve grows somewhat less than the  one for the   $\eta/s =0.08$ case. Also the fluctuations are less pronounced. 

The  PHSD and EPOSir+PHSDe  momentum eccentricities are shown in the middle left   and middle right panels of Fig. \ref{eccp-compare} while the lower  panel 
shows the comparison of the averaged $\epsilon_P$  for all  models.
One can see that the momentum eccentricity in the PHSD model fluctuates much less than the EPOS one, it grows rapidly and saturates after 2 fm/c. The initial EPOSir conditions lead to stronger fluctuations of $\epsilon_P$ in case of EPOSir+PHSDe, however, the PHSD dynamical evolution takes over and leads to a saturation of the $\epsilon_P$ approximately at the same averaged value as for the PHSD.

This comparison shows that the "evolution", i.e. realization of the partonic and hadronic interactions during the expansion,  is more important than the "initial conditions" for the building of pressure.  Indeed, starting with identical initial conditions for  EPOSir+PHSDe and  EPOS, we obtain a saturation of $\epsilon_P$ for  EPOSir+PHSDe  and a monotonic increase of $\epsilon_P$ for the EPOS. But we should always keep in mind that we cannot use directly EPOSi as initial condition for PHSDe, but we need to add the rope procedure ($\to$ EPOSir).

The strongly increasing $\epsilon_P$ has consequences on final observables: in EPOS,  a strong asymmetric flow is associated  to a strong radial flow; 
the latter is visible in the $p_T$ spectra ($dN/dp_T$) at intermediate $p_T$ (1-4 GeV/c), the former in the $p_T$ dependence of $v_2$ (we note that turning off the flow, both $dN/dp_T$ and $v_2$($p_T$) would drop significantly).

\section{Bulk matter observables}

In this section we present and compare the results from the three models - EPOS, PHSD and EPOSir+PHSDe - for the bulk matter observables such as the rapidity and pseudorapidity densities, transverse momentum spectra, transverse mass spectra, and anisotropic flow coefficients for Au+Au collisions at invariant energy $\sqrt{s_{NN}}=200$ GeV. Moreover, we confront our calculations with corresponding experimental data from the BRAHMS, PHENIX, PHOBOS and STAR collaborations.

\subsection{Rapidity and pseudorapidity distributions}

\begin{figure}
\centering
\includegraphics[width=\columnwidth]{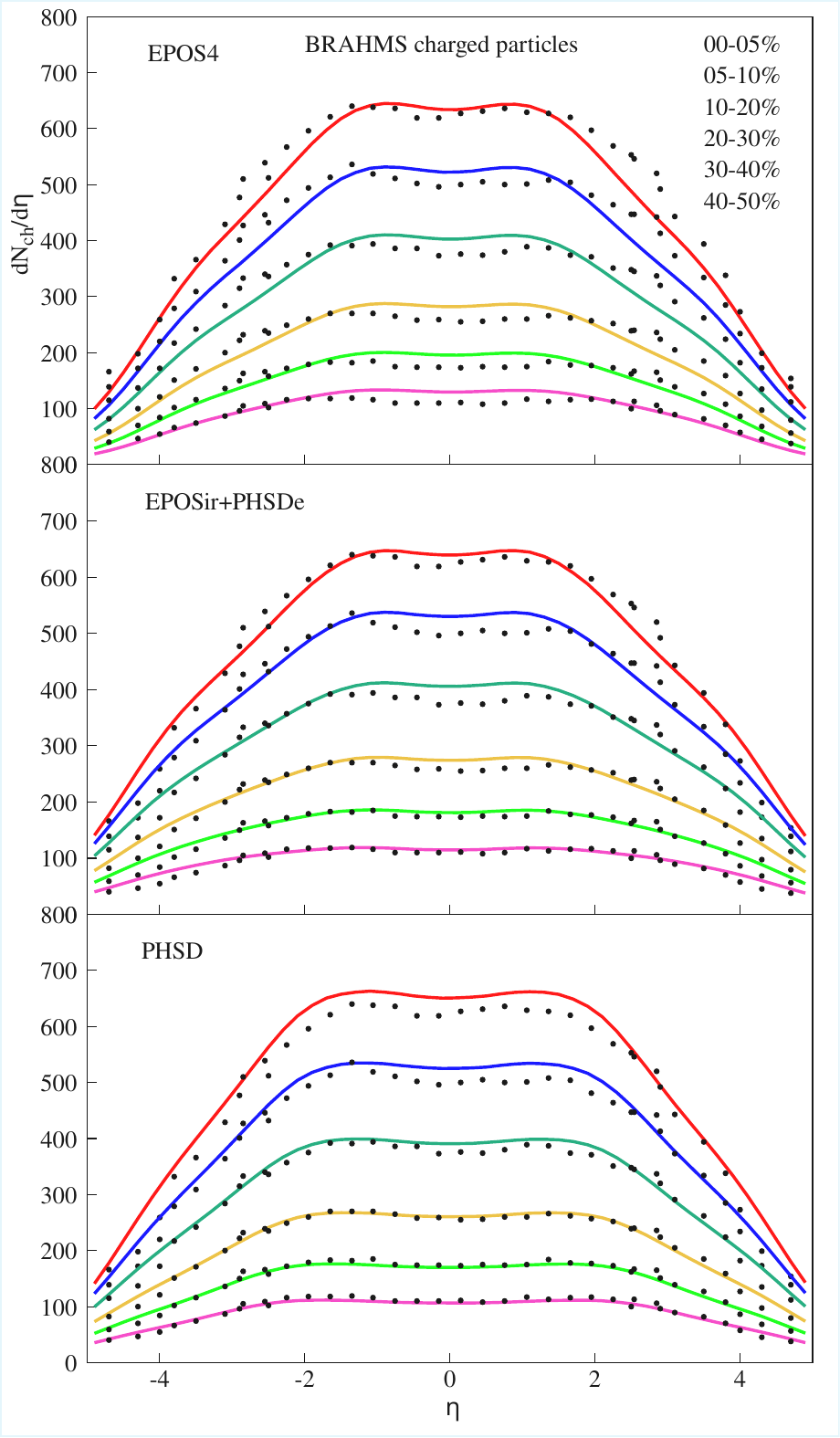}
\caption{Charged particle distribution ($dN_{ch}/d\eta$) as a function of pseudorapidity ($\eta$) for Au+Au collisions at $\sqrt{s_{NN}}=200$ GeV for EPOS (top panel), EPOSir+PHSDe (middle panel), and  PHSD (lower panel) from central to semi-peripheral collisions (from top to bottom: 0-5$\%$, 5-10$\%$, 10-20$\%$, 20-30$\%$, 30-40$\%$, and 40-50$\%$). The experimental data (black dots) are taken from the BRAHMS collaboration \cite{BRAHMS:2001llo}. }
\label{psuedorapidity-3simul}
\end{figure}

We start with showing the  pseudorapidity distributions of charged hadrons.
In Fig. \ref{psuedorapidity-3simul}, we present the EPOS (top panel), EPOSir+PHSDe (middle panel), and  PHSD (lower panel) results for the  pseudorapidity distributions of charged hadrons from most central (0-5$\%$) to semi-peripheral (40-50$\%$) Au+Au collisions at $\sqrt{s_{NN}}=200$ GeV in comparison to the BRAHMS data \cite{BRAHMS:2001llo}. One can see that the PHSD  describes the BRAHMS data very well for all centrality classes with the same shape as the experiment. EPOS approximately reproduces the experimental data, however, the $\eta$-distribution from EPOS is more narrow than the data for all centrality bins.  EPOSir+PHSDe produces slightly more charged particles at mid-pseudorapidity, however, it can well reproduce experimental data for semi-peripheral collisions.

\begin{figure}
\centering
\includegraphics[width=\columnwidth]{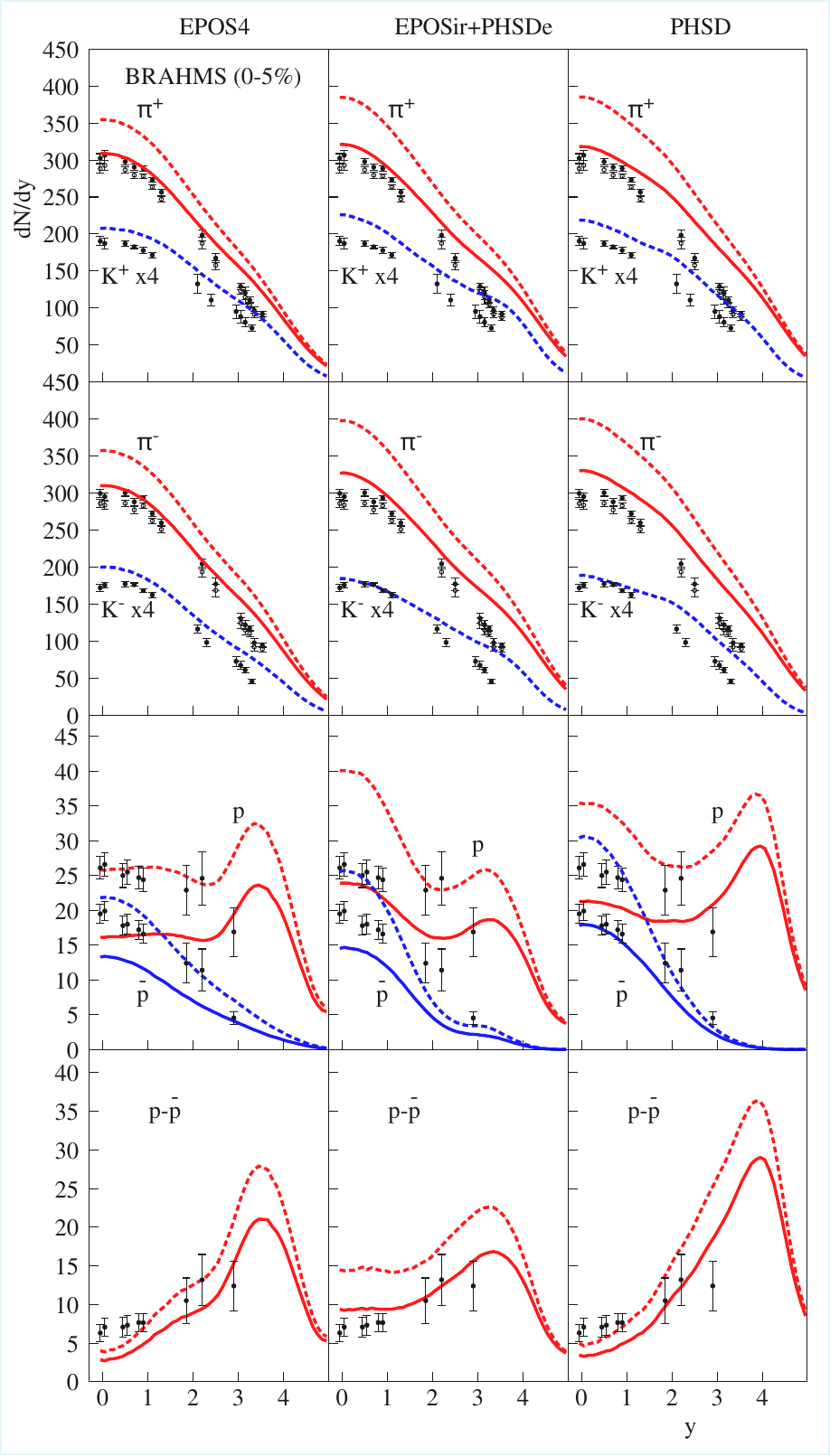}
\caption{Rapidity distributions of (from top to bottom) $\pi^{+}$ (red curves) and $K^{+}$ (blue),  $\pi^{-}$ (red) and $K^{-}$ (blue), $p$ (red) and $\Bar{p}$ (blue), and net protons $p-\Bar{p}$ (red)  for the 5$\%$ most central Au+Au collisions at  $\sqrt{s_{NN}}=200$ GeV, from EPOS (left panel), EPOSir+PHSDe (middle panel), and  PHSD (right panel). The solid lines indicate the results without weak decays, while the dashed lines refer to results containing weak decay products. The kaon yields were multiplied by 4 for clarity. The black dots indicate  the BRAHMS data \cite{BRAHMS:2004dwr, BRAHMS:2003wwg} without weak decay corrections (in case of pions also with).}
\label{rapidity-density}
\end{figure}

In Fig. \ref{rapidity-density}, we present the rapidity densities for identified charged particles ($\pi^{\pm}$ (red curves), $K^{\pm}$ (blue curves), $p$ (red curves), $\bar{p}$ (blue curves), and net proton $p-\bar{p}$ (red curves)) from  EPOS (left panel), EPOSir+PHSDe (middle panel), and  PHSD (right panel) for most central (0-5$\%$) Au+Au collision at the invariant energy $\sqrt{s_{NN}}=200$ GeV.  
The theoretical calculations (integrated over all $p_T$) are compared to the BRAHMS data \cite{BRAHMS:2004dwr, BRAHMS:2003wwg} and shown for two cases:  including the weak decays (dashed lines) or not (solid lines). 
This concerns pions, where weak decays of $K_S$ play an important role, and protons, where $\Lambda$ decays are significant.

The first two rows of Fig. \ref{rapidity-density} represent the $\pi^{\pm}$ (red curves) and $K^{\pm}$ (blue curves) rapidity densities. In all simulations, pions and kaons are a bit above the experimental data.
One can see that the rapidity distributions for light mesons from all three models are slightly broader than the experimental data contrary to the results for the charged particle distributions shown in Fig. \ref{psuedorapidity-3simul} where the model $\eta$-distributions are consistent with the data or even narrower than the data.

The rapidity distributions of  protons (red curves), antiprotons (blue curves), and net-protons are shown in the third and fourth rows of Fig. \ref{rapidity-density}.  Whereas the experimental proton and antiproton rapidity densities decrease with rapidity,  the net-proton density grows. The simulations show a similar trend.
Concerning weak decays, the situation is not clear from the experimental side. The data are not feed-down corrected, but from the detector design, we do not expect weak decay products to contribute considerably. So the data are in reality between "feed-down corrected" and "all feed-down included". If we consider the average of the net-proton results with and without the contribution of weak decays, we find that EPOSir+PHSDe produces results that are closer to the real data than EPOS and  PHSD.
In PHSD, the proton, antiproton, and net-proton distributions with or without weak decay contribution are slightly higher than EPOS and EPOSir+PHSDe in forward rapidity ($3 < y < 5$).

\subsection{Transverse mass spectra}

\begin{figure}
\centering
\includegraphics[width=\columnwidth]{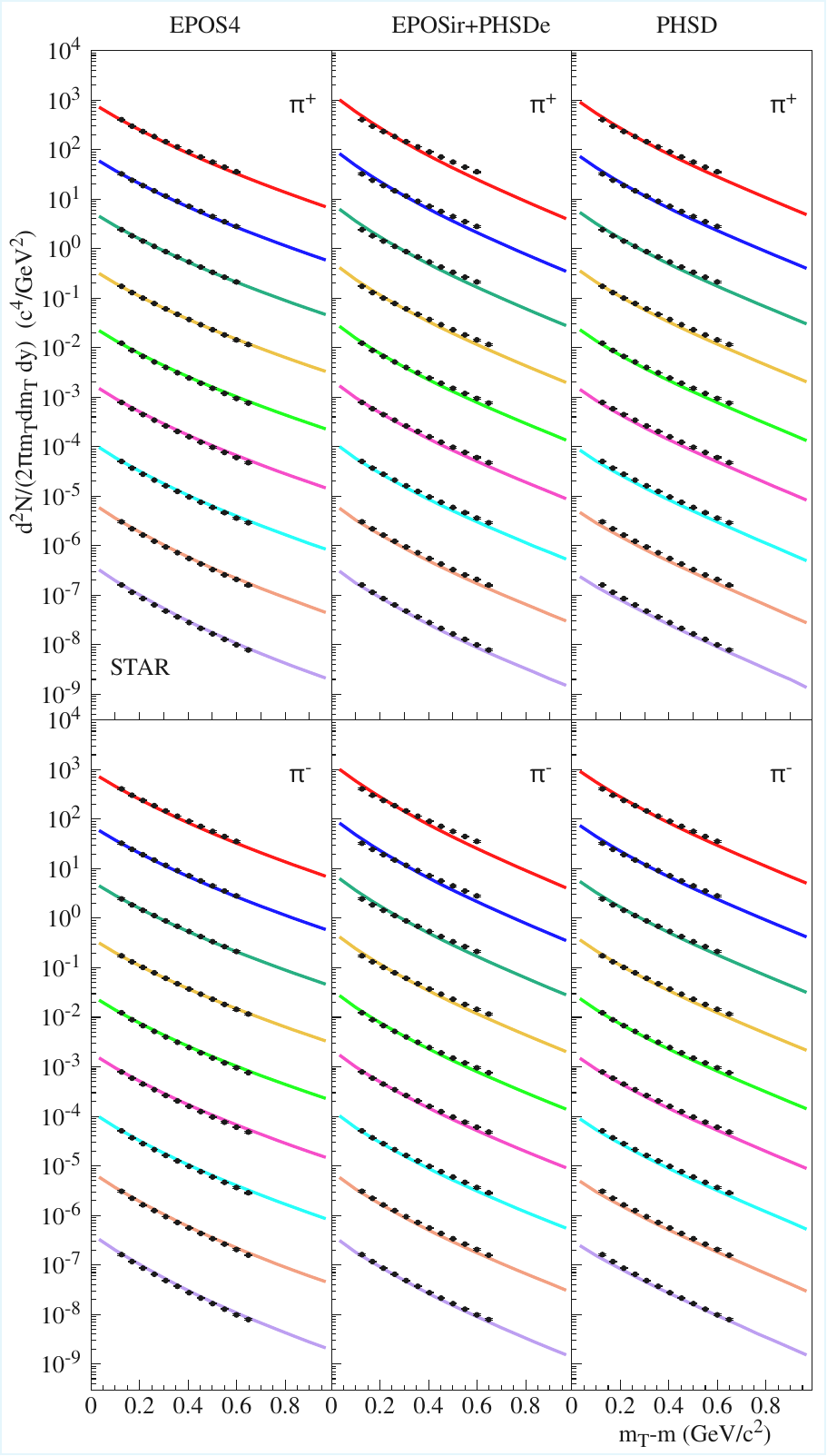}
\caption{Invariant yield as a function of transverse mass for $\pi^{+}$ and $\pi^{-}$  in Au+Au collision at $\sqrt{s_{NN}}=200$ GeV at mid-rapidity ($|y| < 0.5 $) for EPOS (left panel), EPOSir+PHSDe (middle panel) and PHSD (right panel). The $m_T$- spectra are plotted for different centrality bins, from top to bottom: 0-5$\%$, 5-10$\%$, 10-20$\%$, 20-30$\%$, 30-40$\%$, 40-50$\%$, 50-60$\%$, 60-70$\%$, 70-80$\%$, in each plot. The results are compared to the STAR experimental data \cite{STAR:2003jwm} (black dots). All curves and experimental data are scaled by $10^{-n}$ starting from the top curve with $10^{0}$. }
\label{transverse-mass}
\end{figure}
\begin{figure}
\centering
\includegraphics[width=\columnwidth]{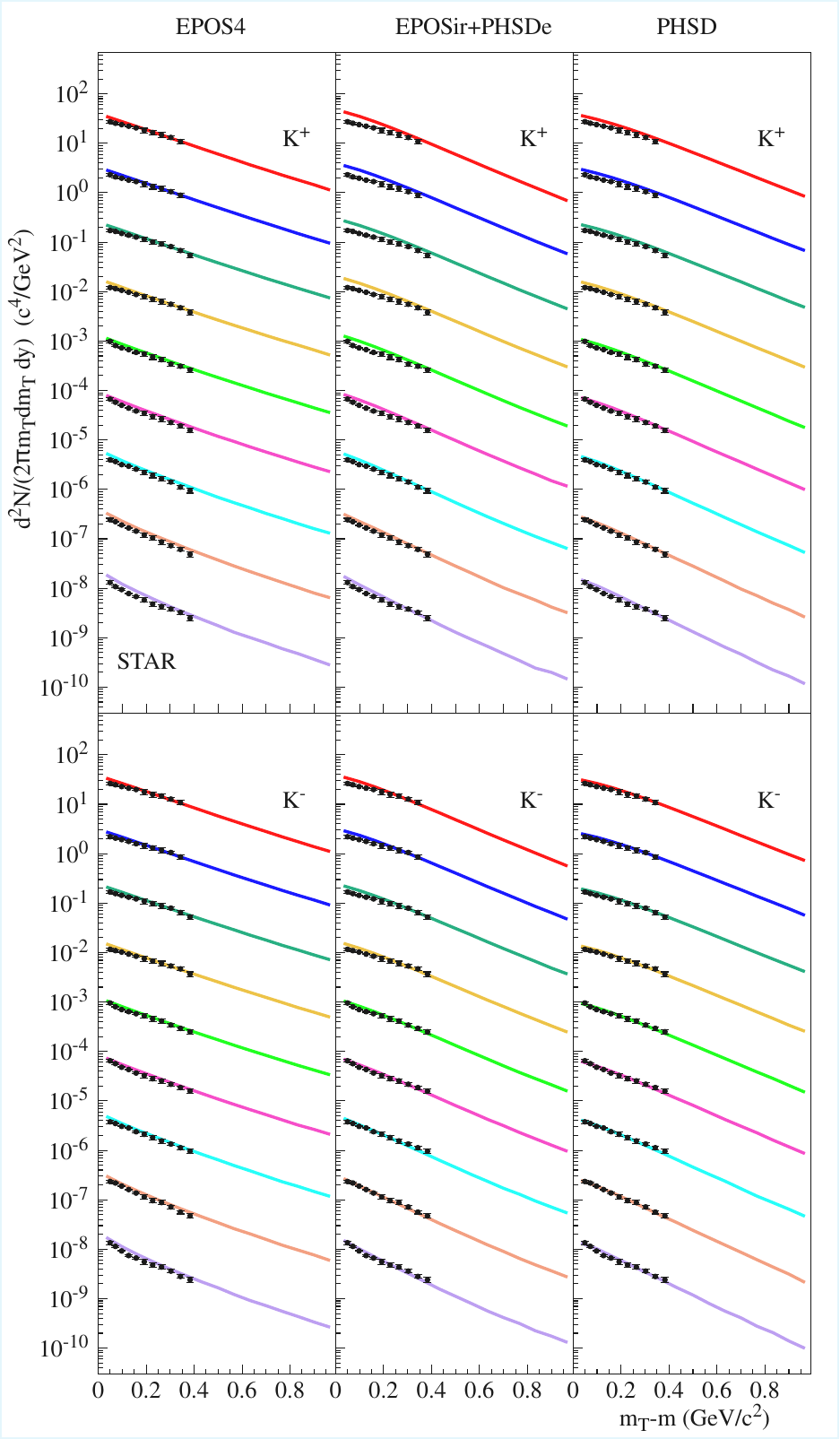}
\caption{Invariant yield as a function of transverse mass for $K^{+}$ and $K^{-}$  in Au+Au collision at $\sqrt{s_{NN}}=200$ GeV at mid-rapidity ($|y| < 0.5 $) for EPOS (left panel), EPOSir+PHSDe (middle panel) and PHSD (right panel). The $m_T$- spectra are plotted for different centrality bins, from top to bottom: 0-5$\%$, 5-10$\%$, 10-20$\%$, 20-30$\%$, 30-40$\%$, 40-50$\%$, 50-60$\%$, 60-70$\%$, 70-80$\%$, in each plot. The results are compared to the STAR experimental data \cite{STAR:2003jwm} (black dots). All curves and experimental data are scaled by $10^{-n}$ starting from the top curve with $10^{0}$. }
\label{transverse-mass1}
\end{figure}
\begin{figure}
\centering
\includegraphics[width=\columnwidth]{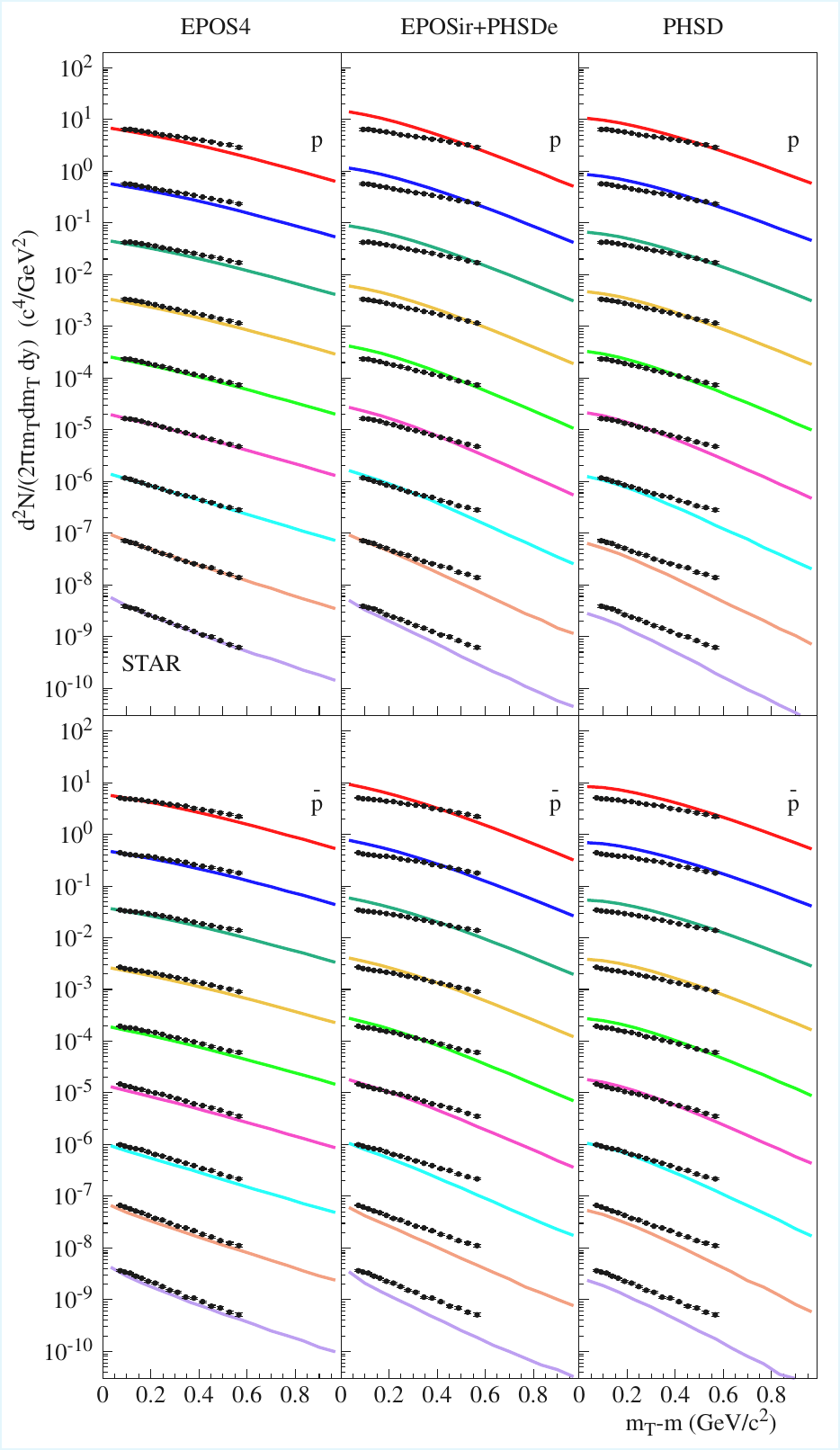}
\caption{Invariant yield as a function of transverse mass for $p$ and $\bar{p}$  in Au+Au collision at $\sqrt{s_{NN}}=200$ GeV at mid-rapidity ($|y| < 0.5 $) for EPOS (left panel), EPOSir+PHSDe (middle panel) and PHSD (right panel). The $m_T$- spectra are plotted for different centrality bins, from top to bottom: 0-5$\%$, 5-10$\%$, 10-20$\%$, 20-30$\%$, 30-40$\%$, 40-50$\%$, 50-60$\%$, 60-70$\%$, 70-80$\%$, in each plot. The results are compared to the STAR experimental data \cite{STAR:2003jwm} (black dots). All curves and experimental data are scaled by $10^{-n}$ starting from the top curve with $10^{0}$. }
\label{transverse-mass2}
\end{figure}

In Figs. \ref{transverse-mass}, \ref{transverse-mass1}, and  \ref{transverse-mass2},
we present the transverse mass ($m_{T}=\sqrt{p_{T}^{2}+m^{2}}$) spectra for charged pions ($\pi^{\pm}$), charged kaons ($K^{\pm}$), as well as protons ($p$) and antiprotons ($\bar{p}$), respectively, for Au+Au collision at $\sqrt{s_{NN}}=200$ GeV for the three models in comparison to STAR data \cite{STAR:2003jwm}. 
We show results for 9 centrality classes, i.e., (from top to bottom for each plot) 0-5$\%$, 5-10$\%$, 10-20$\%$, 20-30$\%$, 30-40$\%$, 40-50$\%$, 50-60$\%$, 60-70$\%$, 70-80$\%$. 

As seen from Figs. \ref{transverse-mass}, \ref{transverse-mass1} all three models can reproduce well the experimental results for all centrality classes in the case of pions ($\pi^{\pm}$) and kaons/antikaons ($K^{\pm}$). The results from EPOSir+PHSDe (middle panel) are quite similar to PHSD (right panel) and EPOS (left panel). 

However, the $m_T$-spectra for the $p$ and $\Bar{p}$ show model deviations:
as follows from Fig. \ref{transverse-mass2}, EPOS reproduces the STAR data for all centrality classes while EPOSir+PHSDe and PHSD show a slightly steeper slope compared to the data and EPOS, respectively.

\subsection{Transverse momentum spectra}\label{pt-spectra-chapter5}

\begin{figure}
\centering
\includegraphics[width=\columnwidth]{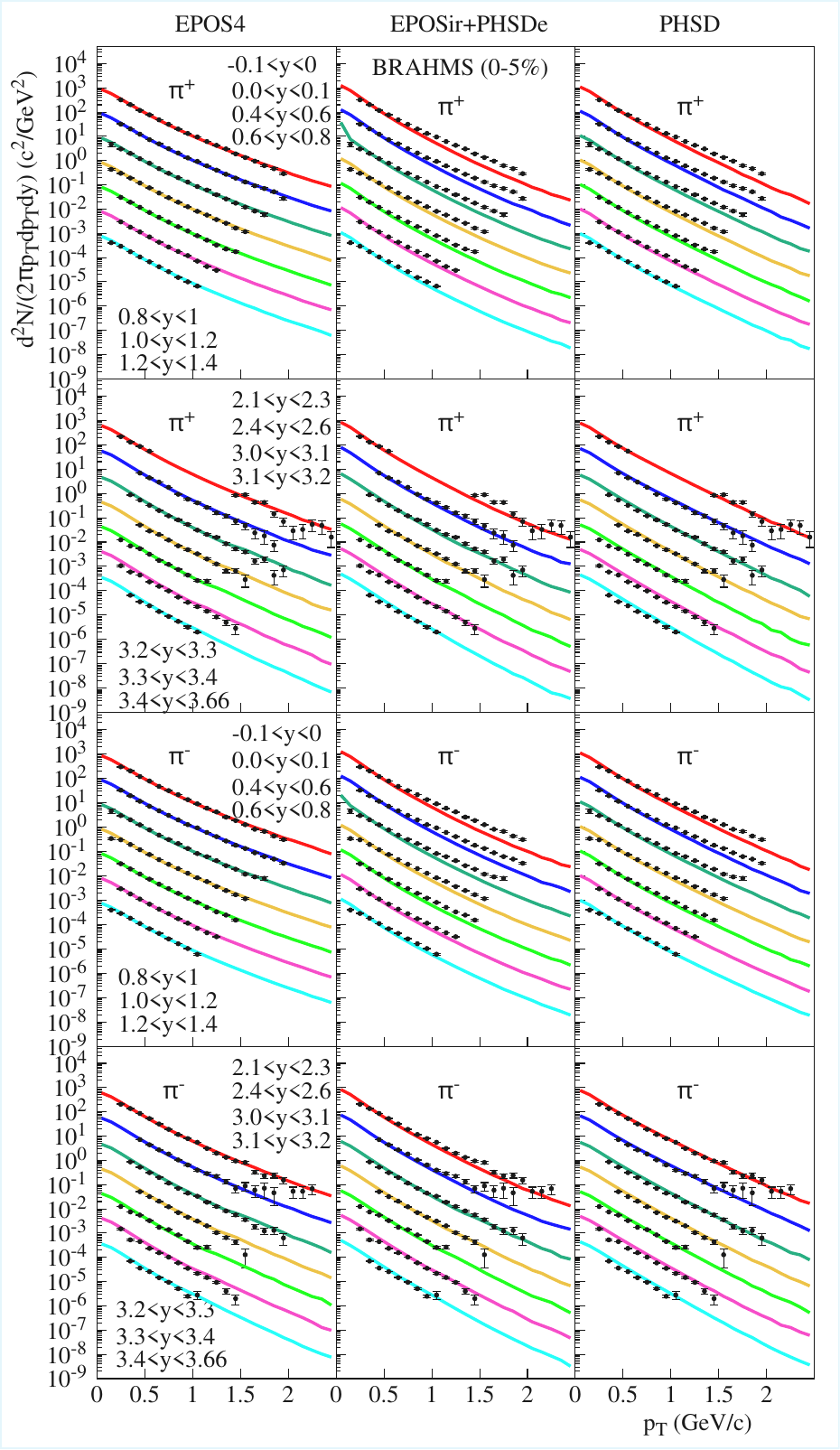}
\caption{Transverse momentum spectra of $\pi^{+}$ and $\pi^{-}$ for selected rapidity intervals (from top to bottom [$-0.1,0$] to [$3.4,3.66$]) for central (0-5$\%$) Au+Au collisions at $\sqrt{s_{NN}}=200$ GeV for EPOS (left panel), EPOSir+PHSDe (middle panel) and  PHSD (right panel). No feed down from weak decays is accounted for. The experimental data are from the BRAHMS collaboration (black dots) \cite{BRAHMS:2004dwr}. Model results are scaled by $10^{-n}$, starting from the top curve with $10^{0}$.} 
\label{pt-brahams2}
\end{figure}

\begin{figure}
\centering
\includegraphics[width=\columnwidth]{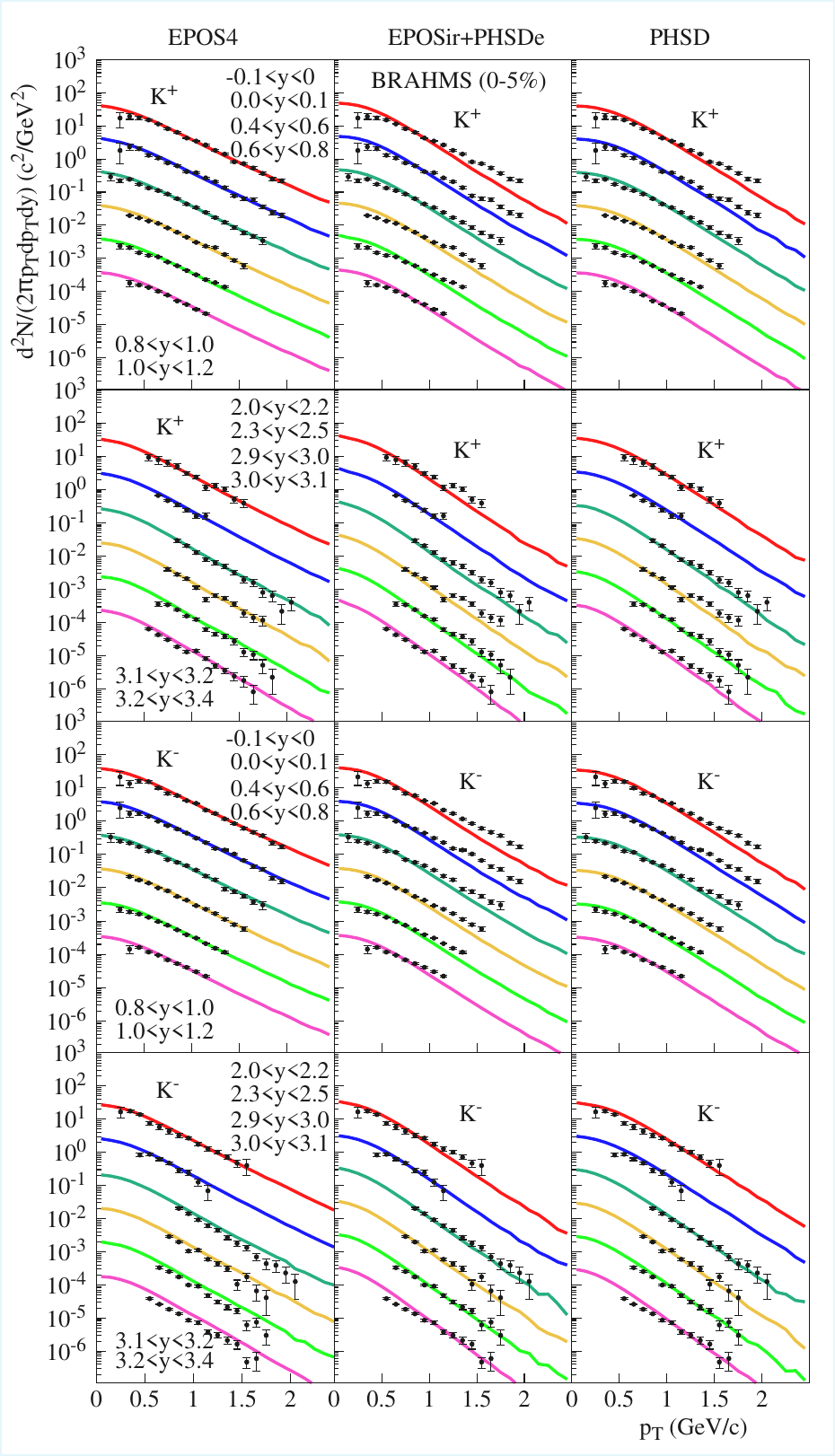}
\caption{Transverse momentum spectra of $K^{+}$ and $K^{-}$ for selected rapidity intervals (from top to bottom [$-0.1,0$] to [$3.2,3.4$]) for central (0-5$\%$) Au+Au collisions at $\sqrt{s_{NN}}=200$ GeV for EPOS (left panel), EPOSir+PHSDe (middle panel) and  PHSD (right panel). Here, feed down from weak decays is accounted for. The experimental data are from the BRAHMS collaboration (black dots) \cite{BRAHMS:2004dwr}. Model results are scaled by $10^{-n}$, starting from the top curve with $10^{0}$.} 
\label{pt-brahams3}
\end{figure}

\begin{figure}
\centering
\includegraphics[width=\columnwidth]{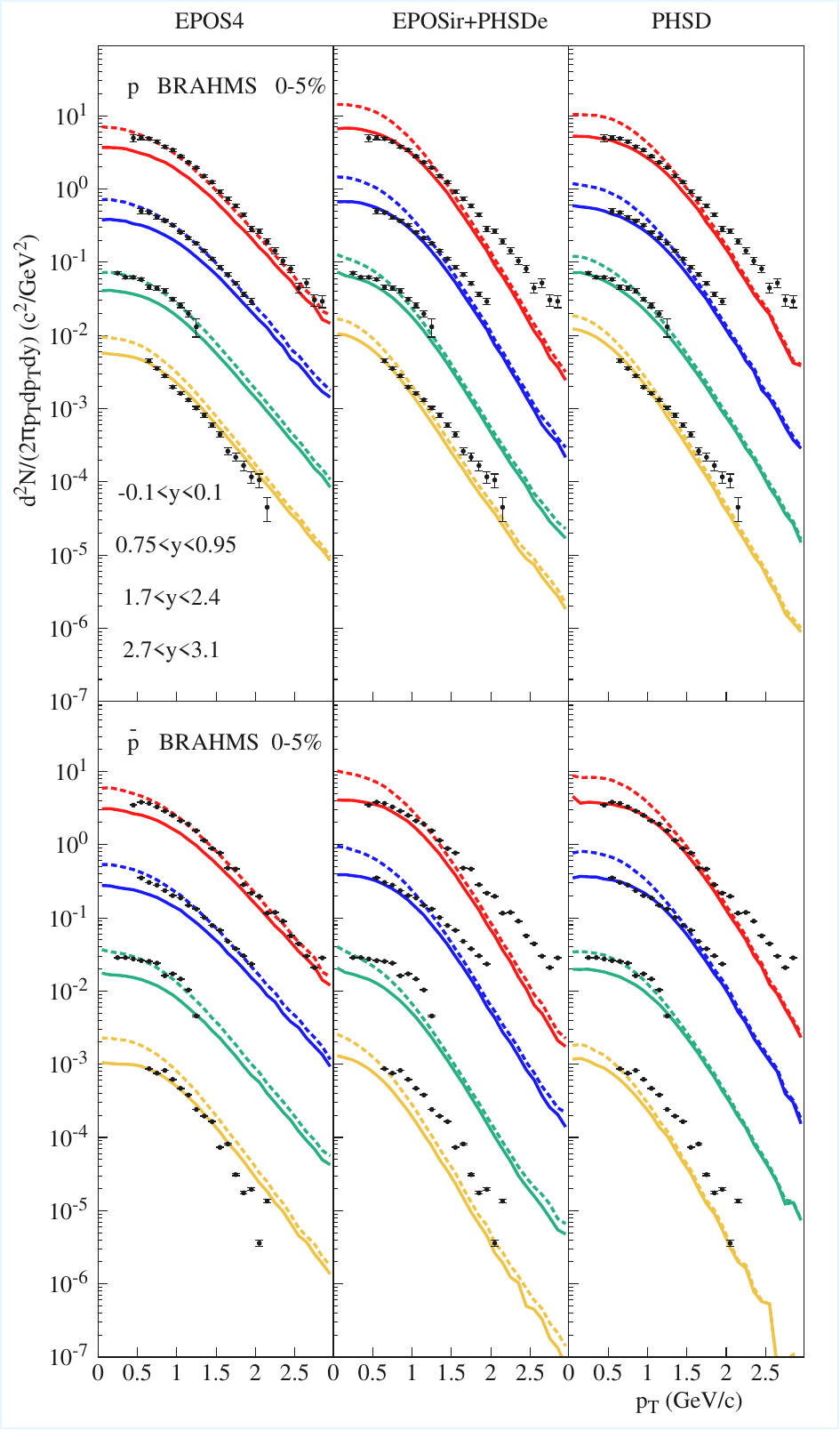}
\caption{Proton ($p$) and anti-proton ($\Bar{p}$) transverse momentum spectra for selected rapidity intervals  (from top to bottom [$-0.1,0.1$] to [$2.7,3.1$]) for central (0-5$\%$) Au+Au collisions at $\sqrt{s_{NN}}=200$ GeV for EPOS (left panel), EPOSir+PHSDe (middle panel) and PHSD (right panel). The dashed and solid lines show the results with and without accounting for weak decays, respectively.
    The experimental data are from the BRAHMS collaboration (black dots) \cite{BRAHMS:2003wwg} and model results are scaled by $10^{-n}$ and $(1/dy)\times 10^{-n}$, respectively, starting from the top curve with $10^{0}$.}
\label{proton-antiproton-pt}
\end{figure}

\begin{figure}
\centering
\includegraphics[width=\columnwidth]{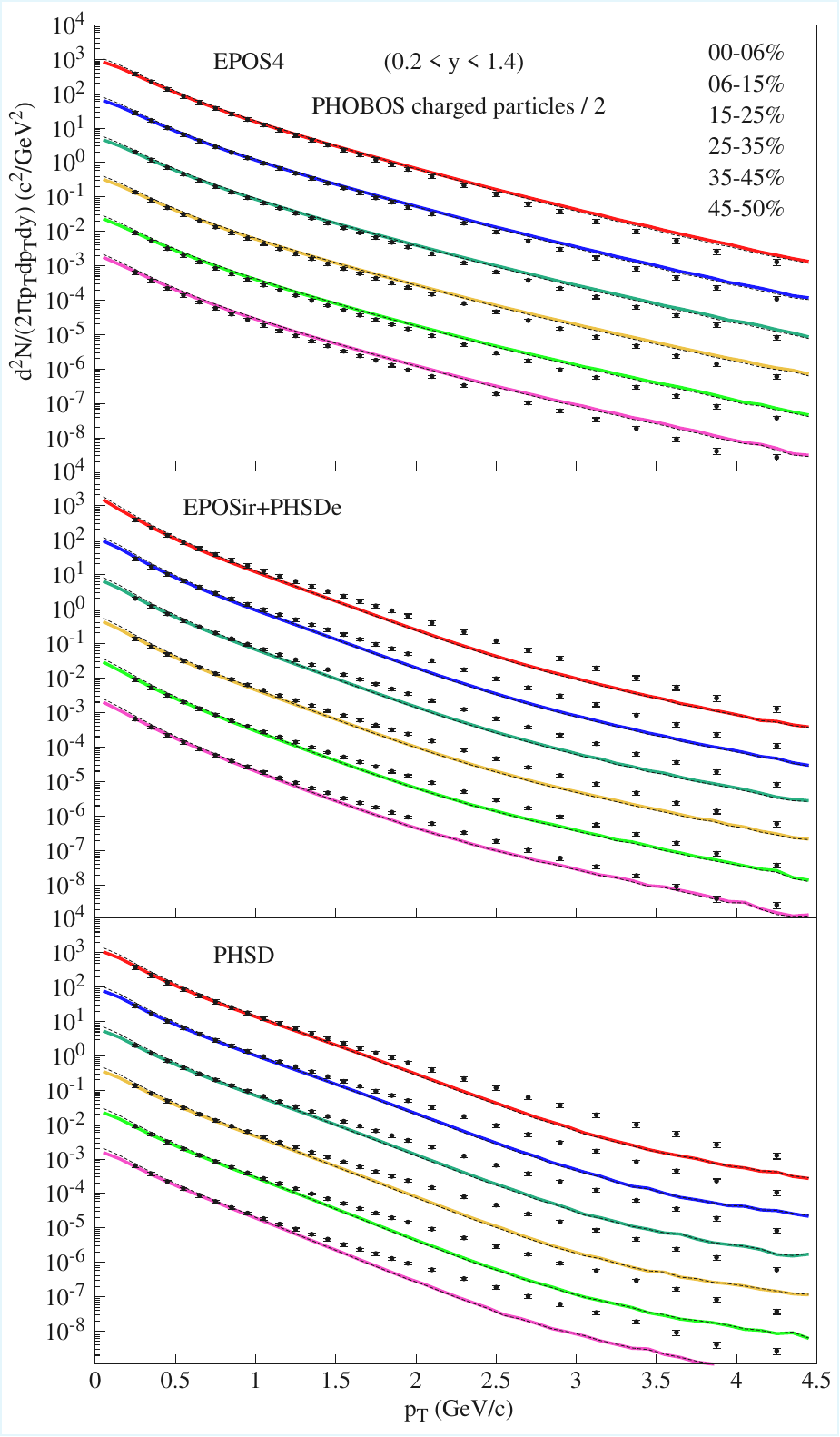}
\caption{Invariant yields for charged hadrons as a function of $p_{T}$ for 6 centrality bins (from top to bottom: 0-6$\%$, 6-15$\%$, 15-25$\%$, 25-35$\%$, 35-45$\%$, and 45-50$\%$) in a rapidity range of $0.2 < y < 1.4$ for  Au+Au collisions at $\sqrt{s_{NN}}=200$ GeV for EPOS (upper panel), EPOSir+PHSDe (middle panel) and PHSD (lower panel). The experimental data from the PHOBOS collaboration (black dots) \cite{PHOBOS:2003wxa}. Simulation's results are scaled by $10^{-n}$ and $(1/dy)\times 10^{-n}$, respectively, starting from the top curve with $10^{0}$. The dashed and solid lines show the results with and without accounting for weak decays, respectively.   }
\label{charged-hadrons-pt}
\end{figure}

\begin{figure}
\centering
\includegraphics[width=\columnwidth]{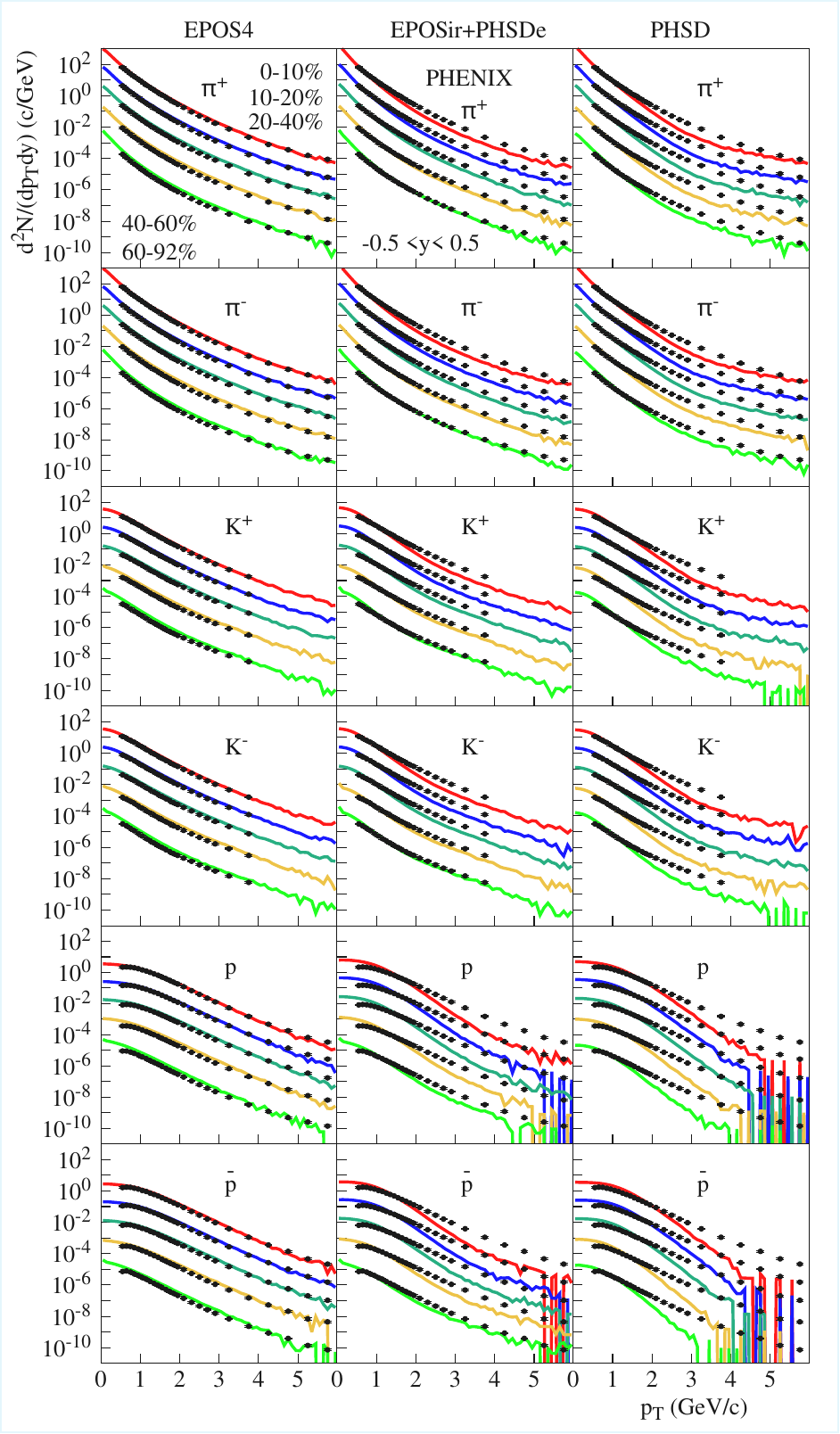}
\caption{Invariant yield of $\pi^+$, $\pi^-$, $K^+$, $K^-$,  $p$, and $\Bar{p}$ as a function of transverse momentum $p_{T}$ at mid-rapidity ($|y| < 0.5$) for Au+Au collisions at  $\sqrt{s_{NN}}=200$ GeV, from central (0-10$\%$) to peripheral (60-92$\%$) collisions (from top to bottom in each plot), for EPOS (left panel), EPOSir+PHSDe (middle panel) and PHSD (right panel). The experimental data are taken from the PHENIX collaboration (black symbols) \cite{PHENIX:2013kod}. All curves and experimental data are scaled by $10^{-n}$ starting from the top curve with $10^{0}$.}
\label{pt-phenix1}
\end{figure}

\begin{figure}
\centering
\includegraphics[width=\columnwidth]{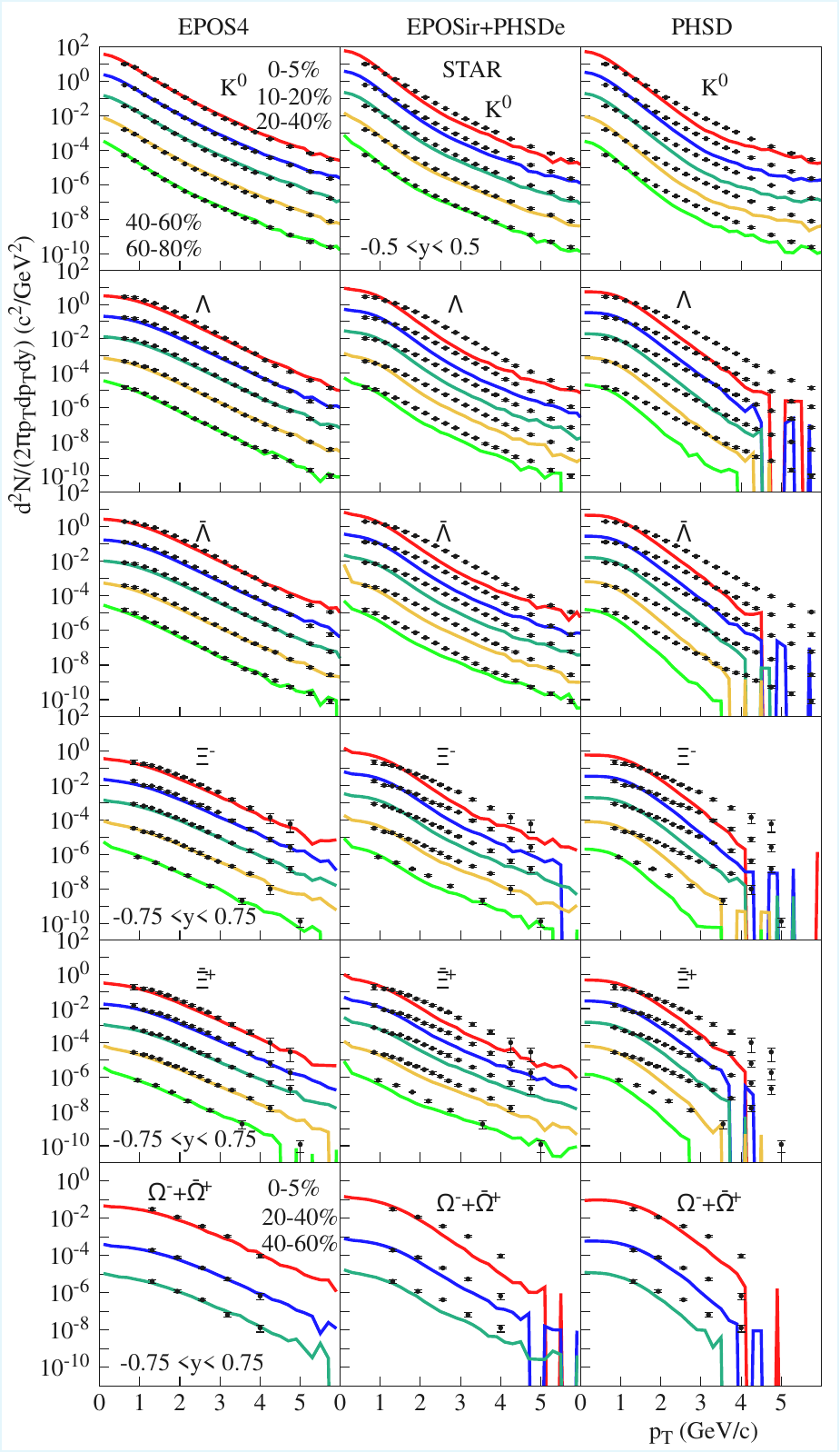}
\caption{Invariant yield of $K_{S}^{0}$,  $\Lambda$, $\Bar{\Lambda}$, $\Xi^{-}$, $\Bar{\Xi}^{+}$, and $\Omega^{-}+\Bar{\Omega}^{+}$ as a function of transverse momentum $p_{T}$ at mid-rapidity for Au+Au collisions at  $\sqrt{s_{NN}}=200$ GeV from  central (0-5$\%$) to peripheral collisions (60-80$\%$) (from top to bottom in each plot) for EPOS (left panel), EPOSir+PHSDe (middle panel), and PHSD (right panel). The experimental data are taken from the STAR collaboration (black dots) \cite{STAR:2006egk}. The experimental data and simulation's results are scaled by $10^{-n}$ and $(1/dy) * 10^{-n}$, respectively, starting from the top curve with $10^{0}$.    }
\label{hyperon-star}
\end{figure}

In this section, in Figs. \ref{pt-brahams2} to \ref{hyperon-star}, we present the invariant yields of identified hadrons as a function of transverse momentum $p_{T}$  and centrality classes for the three models in comparison to experimental data for Au+Au collisions at top RHIC energy. Since the experimental $p_T$ spectra are available up to large $p_T$ values, it extends the comparison of transverse mass distributions from the previous section to a larger momentum space.


In Figs. \ref{pt-brahams2}, \ref{pt-brahams3}, and \ref{proton-antiproton-pt}, we show the rapidity dependence of $p_{T}$ spectra for pions $\pi^{\pm}$, kaons $K^{\pm}$,  protons ($p$), and anti-protons ($\Bar{p}$), for central (0-5$\%$) Au+Au collisions at $\sqrt{s_{NN}}=200$ GeV for EPOS (left panels), EPOSir+PHSDe (middle panels) and PHSD (right panels), in comparison to the BRAHMS experiment \cite{BRAHMS:2004dwr, BRAHMS:2003wwg}.
We note that the $p$ and $\bar p$ spectra are shown with and without accounting for the weak decays since 
the experimental procedures consider weak decay products 'partly'.
One can see that the 
pion and kaon spectra are well described by EPOS for all rapidity bins while the $p_T$- spectra from PHSD and EPOSir+PHSDe are slightly softer for mid-rapidity bins. A similar trend is observed for the $p, \bar p$ spectra: the slope from EPOS is harder than from  PHSD and EPOSir+PHSDe and better in line with the BRAHMS data.

Now we study the centrality dependence of the $p_T$ spectra measured close to mid-rapidity for identified hadrons in Au+Au collisions at $\sqrt{s_{NN}}=200$ GeV.
We start with Fig. \ref{charged-hadrons-pt}, which shows the invariant yields for charged hadrons as a function of $p_{T}$ for 6 centrality bins (from top to bottom: 0-6$\%$, 6-15$\%$, 15-25$\%$, 25-35$\%$, 35-45$\%$, and 45-50$\%$), in a rapidity range of $0.2 < y < 1.4$,  for EPOS (upper panel), EPOSir+PHSDe (middle panel) and PHSD (lower panel). The model results are compared to the experimental data from the PHOBOS collaboration (black dots) \cite{PHOBOS:2003wxa}. One can see that all models reproduce the low part of the spectra very well, however, the large $p_T$ spectra are nicely reproduced only by EPOS while the PHSD and EPOSir+PHSDe underestimate the high $p_T$ data.

A similar trend is obtained for the $p_T$-spectra of identified hadrons, shown in Figs. \ref{pt-phenix1}, \ref{hyperon-star}.
In particular,  Fig. \ref{pt-phenix1} shows the comparison of model calculations for $p_T$ spectra of  $\pi^{\pm}$, $K^{\pm}$, $p$, and $\Bar{p}$  at mid-rapidity ($|y| < 0.5$) from most central (0-10$\%$) to most peripheral (60-92$\%$) collisions, versus  STAR data \cite{STAR:2006egk}. Fig. \ref{hyperon-star} presents the $p_T$ spectra of $K_{S}^{0}$ and strange (anti-)baryons $\Lambda$, $\Bar{\Lambda}$, $\Xi^{-}$, $\Bar{\Xi}^{+}$,  and $\Omega^{-}+\Bar{\Omega^{+}}$, at mid-rapidity, from most central (0-5$\%$) to peripheral collisions (60-80$\%$). One can see the very good reproduction of all spectra by EPOS and the growing deviation from the experimental data with increasing $p_T$ for PHSD and EPOSir+PHSDe.

We will discuss and interpret all these results later, but before doing so, we will have a closer look so some key elements of EPOS4, namely the core-corona procedure and the choice  of the shear viscosity.

\subsection{The role of core-corona and viscosity in EPOS4\label{coco1}}

As explained in detail earlier, in EPOS4, primary interactions lead to the production of prehadrons and a core-corona procedure allows to  separate the core and corona prehadrons. The former constitute the core, the latter are considered to be hadrons. The core is meant to be matter, which evolves according to viscous hydrodynamics. After the hadronization of the fluid, the created hadrons as well as the corona prehadrons (having been promoted to hadrons) may still interact via hadronic scatterings. 

The core-corona procedure was first introduced in Ref. \cite{Werner:2007bf}, and updated in Ref. \cite{Werner:2013tya}.
The current implementation is discussed in detail in Ref. \cite{werner:2023-epos4-micro}.

\begin{figure}
\centering
\includegraphics[width=\columnwidth]{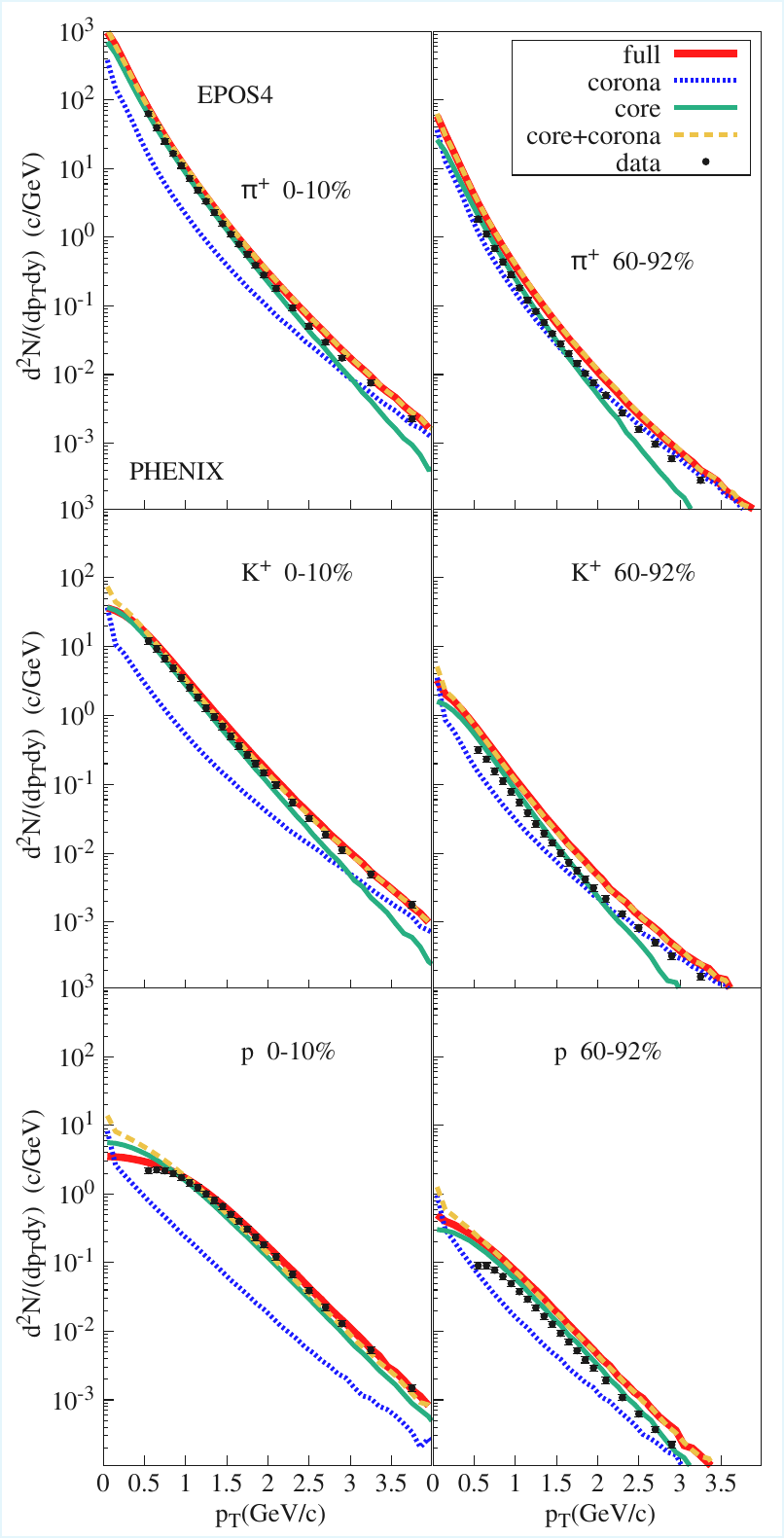}
\caption{Invariant yield of $\pi^{+}$, $K^{+}$, and $p$, as a function of transverse momentum $p_{T}$, at mid-rapidity ($|y| < 0.5$) for Au+Au collisions at $\sqrt{s_{NN}}=200$ GeV for most central (0-10$\%$), and peripheral (60-92$\%$) collisions, for EPOS simulations. The thick red, dotted blue, green, and dashed yellow lines indicate the full (EPOS+hydro +UrQMD), corona, core, and core+corona results. The experimental data are taken from the PHENIX collaboration (black symbols) \cite{PHENIX:2013kod}. }
\label{pt-pikp-coco}
\end{figure}
\begin{figure}
\centering
\includegraphics[width=\columnwidth]{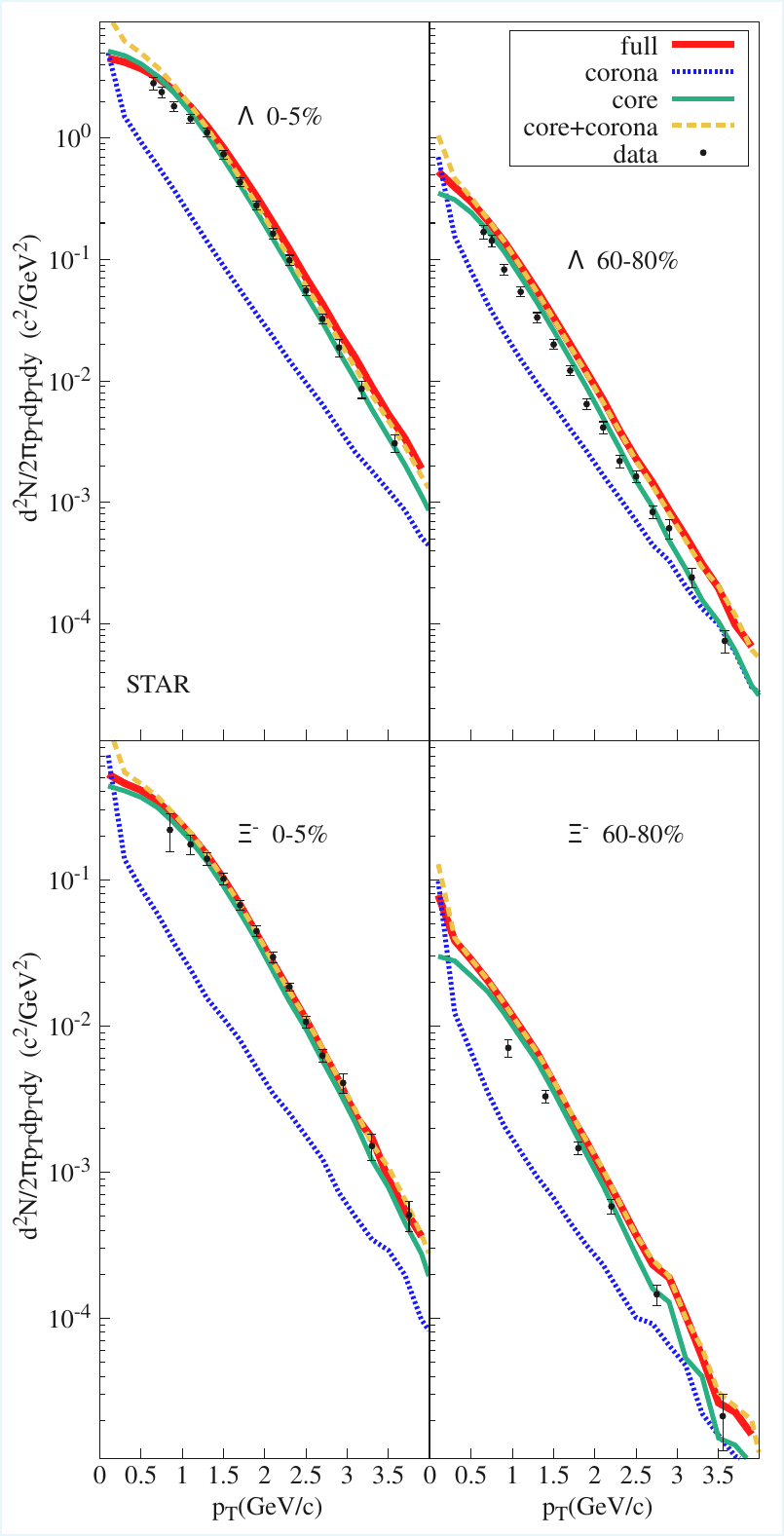}
\caption{Invariant yield of $\Lambda$ and $\Xi^{-}$, as a function of transverse momentum $p_{T}$, at mid-rapidity for Au+Au collisions at  $\sqrt{s_{NN}}=200$ GeV for most central (0-5$\%$), and peripheral (60-80$\%$) collisions, for EPOS4 simulations. The thick red, dotted blue, green, and dashed yellow lines indicate the full (EPOS+hydro +UrQMD), corona, core, and core+corona results. The experimental data are taken from the STAR collaboration (black dots) \cite{STAR:2006egk}. }
\label{pt-lamxi-coco}
\end{figure}

In the following, in Figs. \ref{pt-pikp-coco} and \ref{pt-lamxi-coco}, we show separately  core and corona contributions to hadron production in EPOS4, to better understand the results. We distinguish: 
\begin{description} 
\item [(A)] The "core+corona" contribution: primary interactions (S-matrix approach for parallel scatterings), plus core-corona separation, hydrodynamic evolution, and microcanonical hadronization, but without hadronic rescattering. 

\item [(B)] The "core" contribution: as (A), but considering only core particles. 

\item [(C)] The "corona" contribution: as (A), but considering only corona particles. 

\item [(D)] The "full" EPOS4 scheme: as (A), but in addition hadronic rescattering. 
\end{description}
In cases (A), (B), and (C), we need to exclude the hadronic afterburner, because the latter affects both core and corona particles, so in the full approach, the core and corona contributions are not visible anymore. 

In Fig. \ref{pt-pikp-coco}, the invariant yield of $\pi^{+}$, $K^{+}$, and $p$, is shown as a function of transverse momentum $p_{T}$, at mid-rapidity ($|y| < 0.5$) for Au+Au collisions at $\sqrt{s_{NN}}=200$ GeV for most central (0-10$\%$), and peripheral (60-92$\%$) collisions, for EPOS simulations. The experimental data are taken from the PHENIX collaboration (black symbols) \cite{PHENIX:2013kod}. 
We show separately the different contributions, as discussed in the following.

The green curves refer to the core contribution, i.e., particles produced from the hadronization of the fluid. Most importantly, the latter carry the collective motion (flow velocity) of the fluid, which can be clearly seen at intermediate values of  $p_{T}$ (1-4 GeV/c). Comparing the green core curves for pions, kaons, and protons, one can see that the form changes dramatically: if goes from convex (for pions) to concave (for protons). The important property is the mass of the particles. Since flow is given in terms of velocity, the effect on the transverse momentum is bigger for heavy compared to light particles,  because we have "momentum = mass times velocity". So we get a very strong flow effect for protons, compared to pions. Although pions are much more frequent than protons, the yields are comparable at   $p_{T}$ values of 3-4 GeV/c.

The dotted blue curves refer to the corona contribution, i.e., particles which did not take part in the collective hydrodynamic evolution, but having been produced directly from string decays. If one would multiply the kaon curve by two and the proton one by four, then the three curves  are very close, apart of a bigger yield  for pions at small $p_{T}$ (resonance production). So at intermediate values of   $p_{T}$, it is clearly the flow, visible in the core contribution, which makes a difference between light and heavy particles. 

The dashed yellow curves indicate core+corona results, i.e. the sum of the core and the corona curves. All this (core-corona / core+corona) refers to simulations without hadronic rescattering. Including the latter, on gets the "full" results, shown as thick red lines (the final result). There is a sizable effect (difference between red and yellow) for protons, but anyway small compared to the flow effect.
Comparing the left and the right column in Fig. \ref{pt-pikp-coco}, we see that the core effect is much smaller in peripheral collisions (as expected).  

In Fig. \ref{pt-lamxi-coco}, we show the different contributions for hyperons, which confirms the above discussion, saying that there is a core effect (increased yields at intermediate values of  $p_{T}$ due to flow), which increases with particle mass.
Finally, we note that the "crossing point" of the green and dotted blue curves (from where on the "hard" corona dominates) is at around 3 GeV/c for pions in central collisions, but gets to much larger values with increasing particle mass.

\begin{figure}
\centering
\includegraphics[width=\columnwidth]{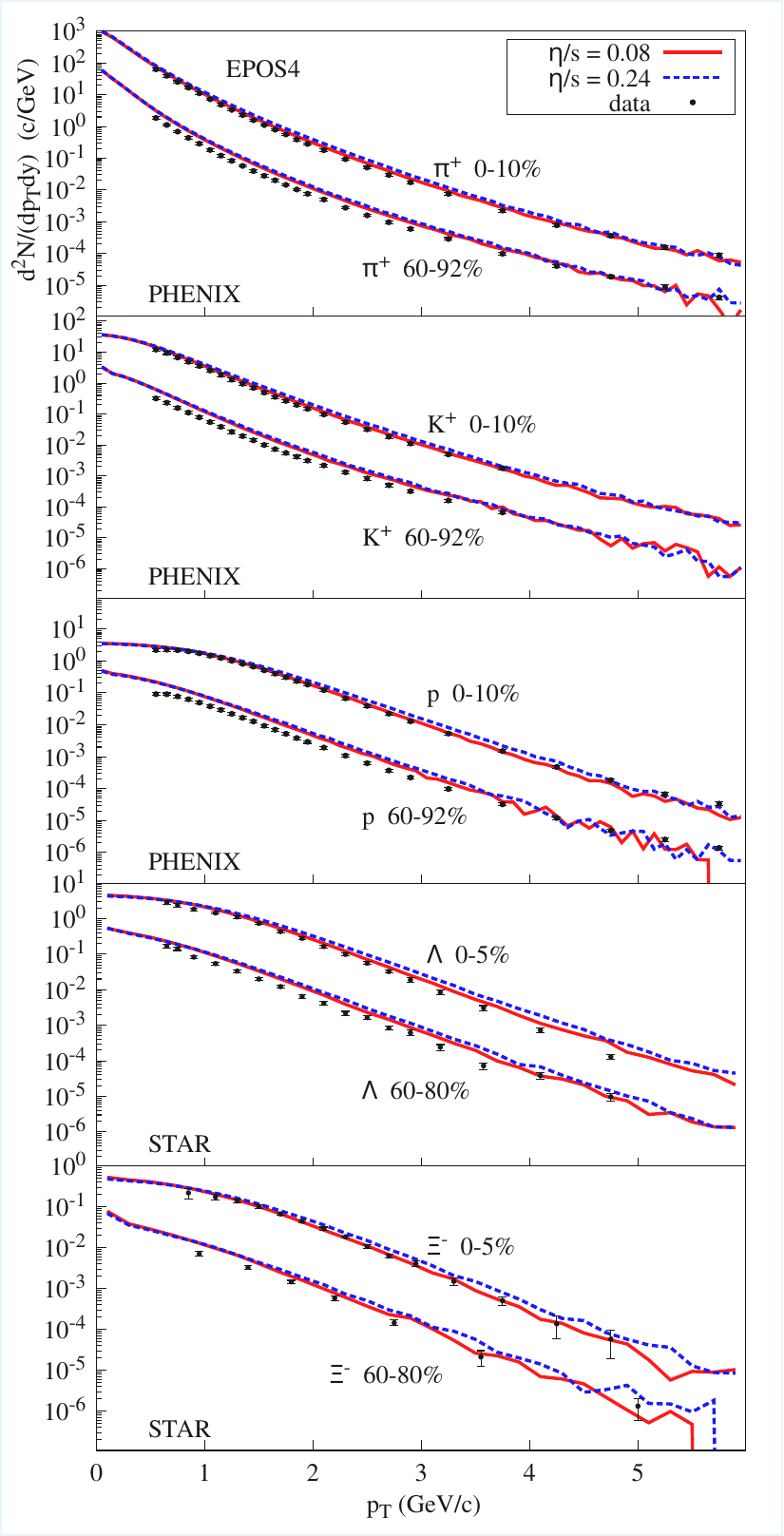}
\caption{Invariant yield of $\pi^+$, $K^+$, $p$, $\Lambda$, and $\Xi^-$, as a function of transverse momentum $p_{T}$, at mid-rapidity, for Au+Au collisions at  $\sqrt{s_{NN}}=200$ GeV, for central and peripheral collisions, for EPOS simulations. The solid red  curves refer to $\eta/s$ of 0.08, the dashed blue ones to 0.24. The experimental data (black dots) are taken from the PHENIX collaboration \cite{PHENIX:2013kod} and from the STAR collaboration \cite{STAR:2006egk}.}
\label{visc1}
\end{figure}
\begin{figure}
\centering
\includegraphics[width=\columnwidth]{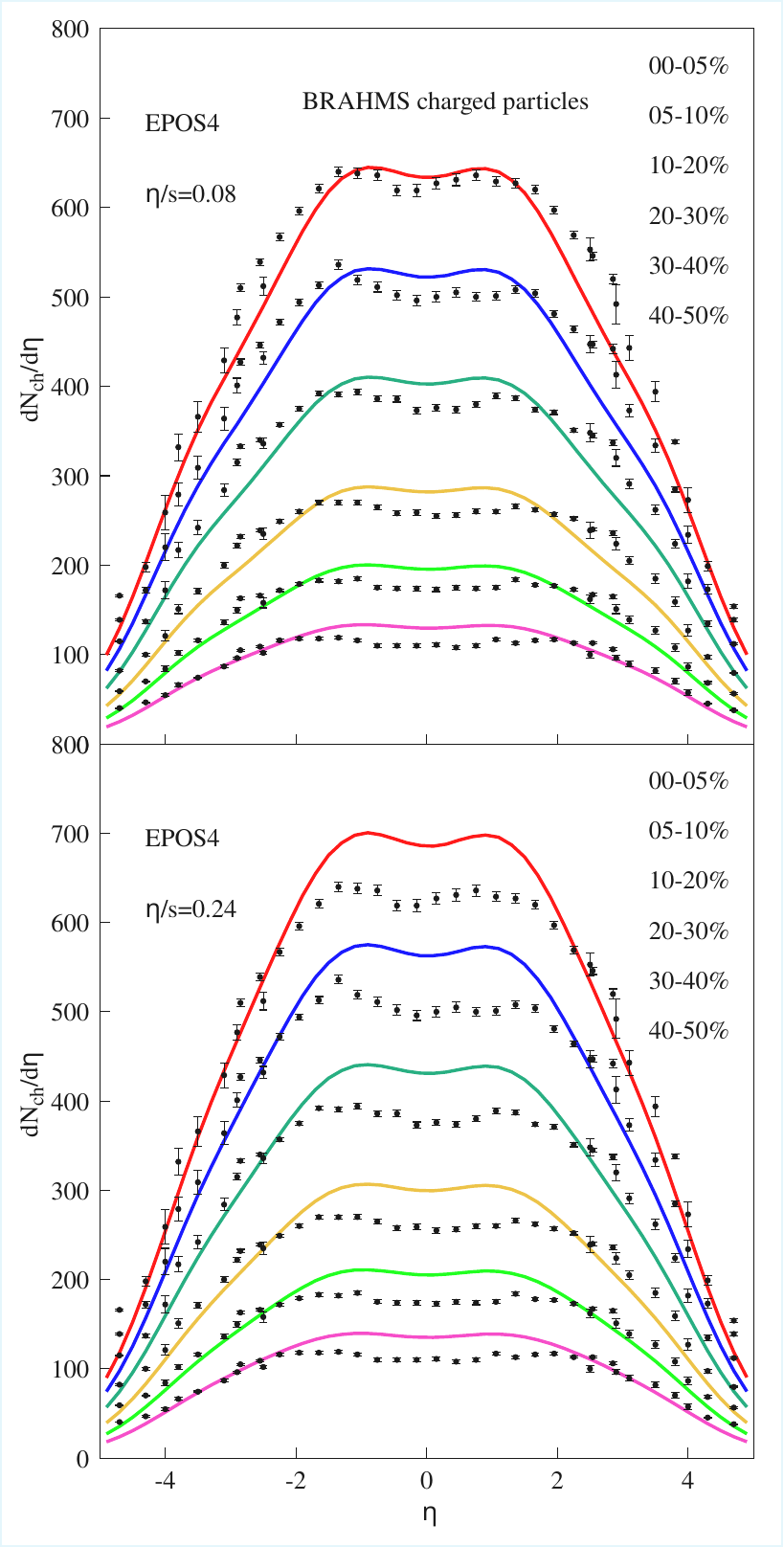}
\caption{Charged particle distribution ($dN_{ch}/d\eta$), as a function of pseudorapidity ($\eta$), for Au+Au collisions at $\sqrt{s_{NN}}=200$ GeV, for EPOS simulations, from central to semi-peripheral collisions (from top to bottom: 0-5$\%$, 5-10$\%$, 10-20$\%$, 20-30$\%$, 30-40$\%$, and 40-50$\%$). The upper panel refers to $\eta/s$ of 0.08, the lower to 0.24. The experimental data (black dots) are taken from the BRAHMS collaboration \cite{BRAHMS:2001llo}. }
\label{visc2}
\end{figure}

In Figs. \ref{visc1} and \ref{visc2}, we investigate another important issue, namely the dependence of EPOS4 results on the shear viscosity. In Fig. \ref{visc1}, we show $p_{T}$ spectra for different particle species for Au+Au collisions at  $\sqrt{s_{NN}}=200$ GeV, for central and peripheral collisions. We show EPOS4 results, using the standard value of o $\eta/s$ of 0.08 (solid red lines) and compare it to results obtained for  $\eta/s$ of 0.24 (dashed blue lines). The latter are a little bit harder, but the difference is small.
Fig. \ref{visc2} shows, that the pseudorapidity distributions get a little bit higher, for $\eta/s=0.24$ compared to 0.08.

\subsection{Messages from the spectra comparison}

The obtained results for the transverse momentum spectra can be interpreted as follows: 
\begin{itemize}
\item
the fact that EPOS and EPOSir+PHSDe, which start with identical initial conditions, but follow a different dynamical evolution, end up with different transverse spectra,
as well as  
\item
the fact that EPOSir+PHSDe and PHSD, which start with completely different initial conditions, but follow approximately a similar dynamical evolution, end up with rather similar spectra,  
\end{itemize}
indicate the dominant role of the dynamical evolution versus the initial conditions for the bulk observables such as transverse momentum/mass spectra. The transverse  momentum spectra at intermediate values of $p_T$ (1-5 GeV/c) are especially sensitive to this balance, i.e.,
\begin{itemize}
\item 
EPOS starts with strongly fluctuating initial conditions in space, transforming quickly into large (compared to PHSD) flow (and anisotropic flow, cf. Fig. \ref{eccp-compare}). This is due to EPOS' core-corona separation, converting the core part into a hydrodynamically evolving matter, assuming very early local thermalization.
The hydrodynamics in EPOS produces at very early times a very strong radial flow, which leads to an enhancement of particle production at intermediate values of $p_T$, in particular  for heavy particles, as discussed in the preceding section. 

\item In PHSD,
the initial conditions, build by primary high energy $NN$ interactions -- realized via the LUND string model -- are less fluctuating than in EPOS since in PHSD the interactions are treated in a probabilistic manner  as independent sequences of collisions in time defined by Monte-Carlo according to the total $NN$ cross sections. Contrary, in EPOS the initial collisions are considered as instantaneous multiple scatterings,  modeled in terms of a summation of amplitudes within S-matrix theory, which allows to account for the interferences of amplitudes.
The treatment of the QGP in the PHSD  provides a good description of the "thermal part" (low   $p_T$) of the spectra, but leads to softer spectra at intermediate values of $p_T$ compared to EPOS as well as to the data. 
The latter can be attributed partially to the modeling of hadronization in PHSD which favors the combination of low $p_T$ partons according to the covariant transition rate \cite{Cassing:2008sv}.
\\
\item EPOSir+PHSDe is based on the EPOS initial conditions, but follows the PHSD dynamical evolutions. There is no  assumption of equilibration. Contrary to the hydrodynamic evolution, the PHSD evolution does not transform the intial space fluctuatins (with local high density "hot spots") into large flow, 
and leads to results which are much closer to PHSD than to EPOS proving the dominant influence of the dynamics.
This finding is in line with an early PHSD study which shows that the size of the initial fluctuations doesn't influence much the final spectra and harmonic flow results \cite{Konchakovski:2014fya}. The latter we will check in the next section.
\end{itemize}

\section{Anisotropic flow}

The study of the azimuthal distribution of particle production provides important information about the system's space-time evolution. 
The anisotropies in the initial geometry of the collisions may be transformed via pressure gradients to the anisotropies in the final state momentum distribution, the latter being accessible experimentally.
The discovery of a large azimuthal anisotropic flow of hadrons at RHIC \cite{STAR:2000ekf} provided one of the first strong signals for dense partonic matter formation in ultra-relativistic nucleus-nucleus collisions, 
producing hydrodynamic flow, investigated also in EPOS \cite{Werner:2010aa}
and PHSD  \cite{Konchakovski:2011qa}.

\subsection{Results for flow harmonics}

\begin{figure}
\centering
\includegraphics[width=\columnwidth]{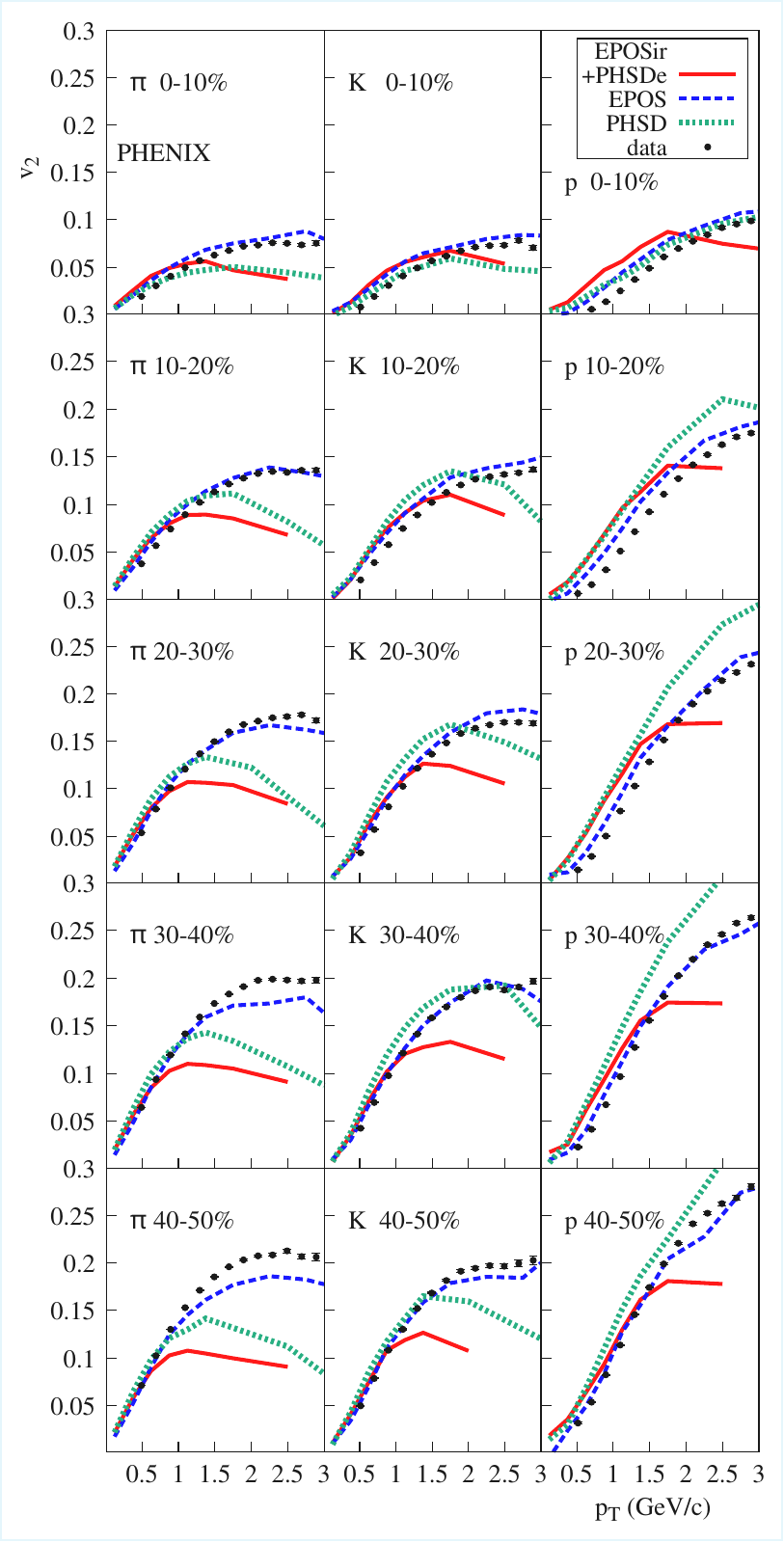}
\caption{Differential elliptic flow ($v_{2}$) of $\pi$, $K$, and $p$, as a function of transverse momentum $p_{T}$,  at mid-rapidity, for Au+Au collisions at $\sqrt{s_{NN}}=200$ GeV, for different centrality bins. Dashed blue, solid red, and dotted green lines indicate the EPOS, EPOSir+PHSDe and  PHSD results, while the black dots show the PHENIX experimental data \cite{PHENIX:2014uik}, respectively. }
\label{v2-pt-RNX}
\end{figure}
\begin{figure}
\centering
\includegraphics[width=\columnwidth]{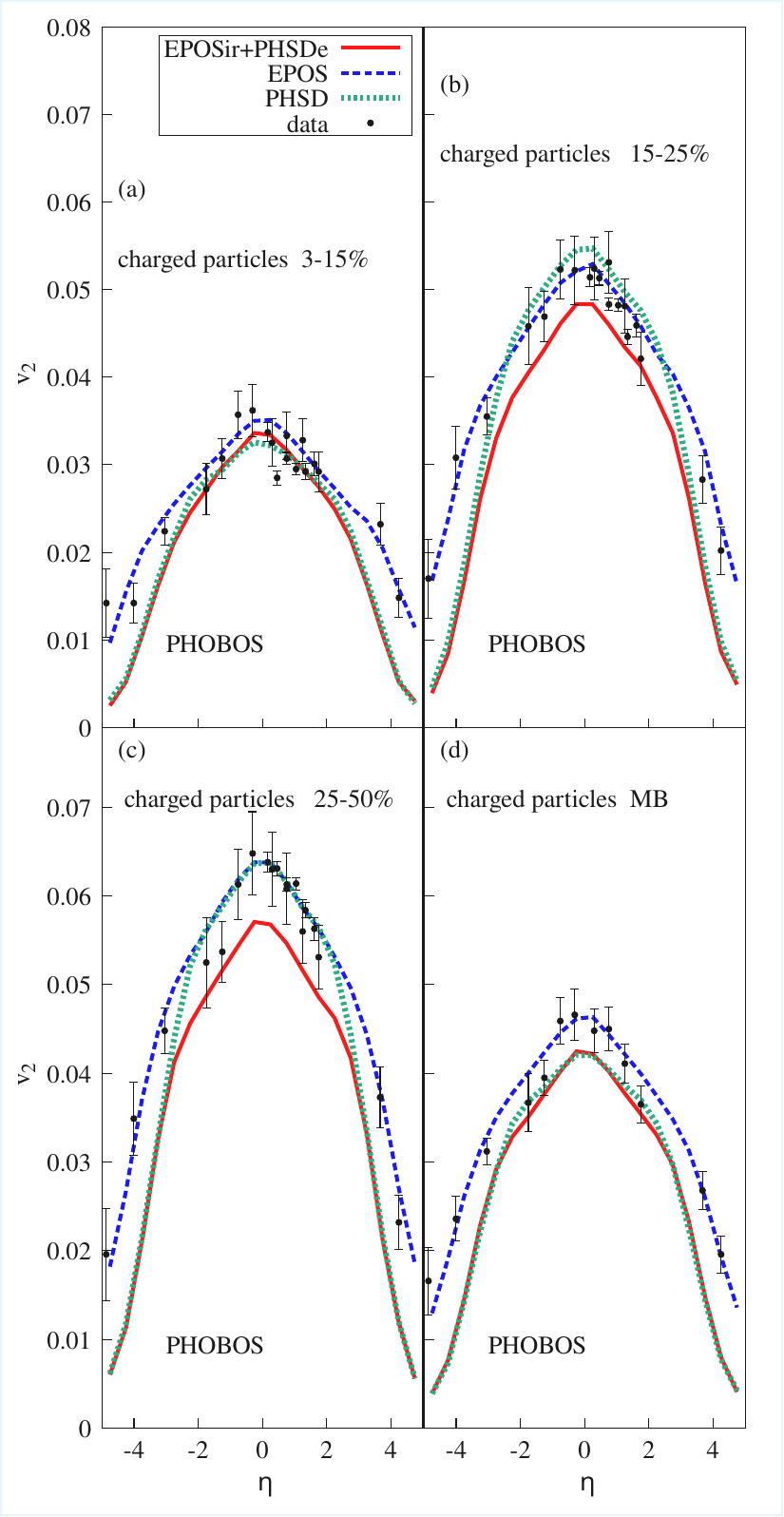}
\caption{Elliptic flow as a function of pseudorapidity ($v_{2}(\eta)$) for charged hadrons in Au+Au collisions at $\sqrt{s_{NN}}=200$ GeV, for different centrality classes, i.e., (a): 3-15$\%$,  (b): 15-25$\%$, (c): 25-50$\%$, and (d): minimum bias (MB). Dashed blue, solid red, and dotted green lines indicate the EPOS, EPOSir+PHSDe and PHSD results, while the black dots show the PHOBOS experimental data \cite{PHOBOS:2004vcu}, respectively.}
    \label{v2-eta-phobos04}
\end{figure}
The Event Plane (EP) approach \cite{Voloshin:2007pc, Poskanzer:1998yz} has been used in the RHIC experiments we are going to compare with, and 
in our analysis. One uses the Fourier series, i.e.,
\begin{equation} 
\frac{d^{3}N}{d^2p_{T}d\eta}  \propto  \Big \{ 
    1+2 \sum_{n=1}^{\infty} v_{n}(p_{T},\eta)\cos[n(\phi - \Psi_{EP})]  \Big \},
\label{fourier-eventplane}
\end{equation}
where $p_{T}$ is the transverse momentum of a particle, $\phi$ the azimuthal angle, $\eta$ the pseudorapidity, and $\Psi_{EP}$ the event plane angle.
Elliptic flow is represented by $v_2=$ \\
$\left< \cos{[ n(\phi - \Psi_{EP})]}\right>$, where the angular brackets denote an average over the particles and events.  

In Figs. \ref{v2-pt-RNX} and \ref{v2-eta-phobos04}, we present the results for the flow harmonics  $v_2$ of the three models, EPOS (dashed blue curves), EPOSir+PHSDe (solid red curves), and PHSD (dotted green curves), and compare them to experimental data.
In Fig. \ref{v2-pt-RNX}, we show the differential flow $v_{2}(p_T)$  of  $\pi$, $K$, and $p$ using the EP method for Au+Au at $\sqrt{s_{NN}}=200$ GeV for different centralities for EPOS, EPOSir+PHSDe, and PHSD. In our analysis, we applied the cut $|\eta| <$  0.35 as in the PHENIX experiment \cite{PHENIX:2014uik}.
EPOS provides a  good description of elliptic flow $v_{2}$  for $\pi$, $K$, and $p$ for the whole $p_T$ range. PHSD describes reasonably well the proton $v_2$ for all centralities as well as for kaons up to 2 GeV/c, while the pions $v_2$ deviate from the data already at $p_T >1$ GeV/c. The elliptic flow from EPOSir+PHSDe is closer to the PHSD results and even slightly lower for pions and kaons and high momentum protons.  The lowering of the $v_2$ in EPOSir+PHSDe relative to the PHSD is in line with the time evolution of momentum eccentricity $\epsilon_P$ in Fig. \ref{eccp-compare}, which shows that $\epsilon_P$ of EPOSir+PHSDe develops slower than the PHSD one and ends with slightly lower averaged value. For both, EPOSir+PHSDe and PHSD,  $\epsilon_P$ grow much less with time than the EPOS curve.
 
In Fig. \ref{v2-eta-phobos04},  we show the elliptic flow $v_2(\eta)$ as a function of pseudorapidity for charged hadrons from central to semi-peripheral as well as for minimum bias Au+Au collisions at $\sqrt{s_{NN}}=200$ GeV for the three models in comparison to the experimental data from the PHOBOS collaboration \cite{PHOBOS:2004vcu}. The $v_2$ from all models and data decreases towards central collisions. As follows from the figure, PHSD provides a good description of the experimental data at mid-presudorapidity, in spite the fact that PHSD underestimates $v_2(p_T)$ for large $p_T$. That is due to the high abundance of hadrons with low $p_T$ which dominantly contribute to  $v_2(\eta)$  integrated over all $p_T$. EPOS provides as well a good description of the data.
The EPOSir+PHSDe flow is slightly lower. 
The $v_2(\eta)$ distribution from EPOS is much wider than those from other two models and has a non-Gaussian shape. 
Thus, $v_2(\eta)$ indicates how the pressure is redistributed over rapidity in the system.

Before discussing and interpreting these results, we will again first discuss some issues related to EPOS4.

\subsection{The role of core-corona and viscosity in EPOS4\label{coco}} 

\begin{figure}
\centering
\includegraphics[width=\columnwidth]{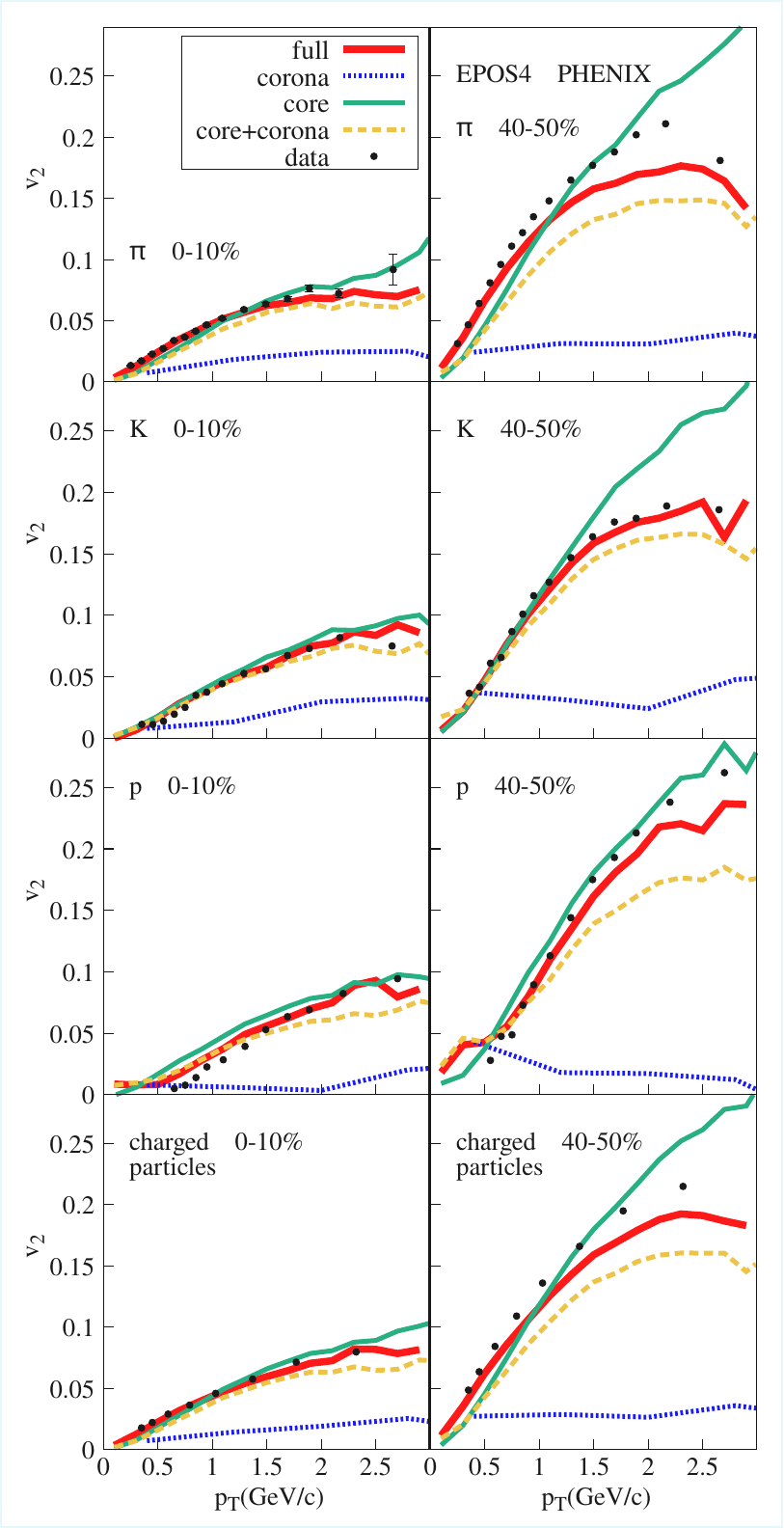}
\caption{Differential elliptic flow ($v_{2}$) of $\pi$, $K$, $p$, and charged particles, as a function of transverse momentum $p_{T}$, at mid-rapidity, for Au+Au collisions at $\sqrt{s_{NN}}=200$ GeV, for central (0-5$\%$) and mid-peripheral (40-50$\%$) collisions, for EPOS simulations. Thick red, dotted blue, green, and dashed yellow lines indicate the full (EPOS+hydro+UrQMD), corona, core, and core+corona results, while the black dots show the PHENIX experimental data \cite{PHENIX:2014uik}, respectively.}
\label{v2pt-coco}
\end{figure}

\begin{figure}
\centering
\includegraphics[width=\columnwidth]{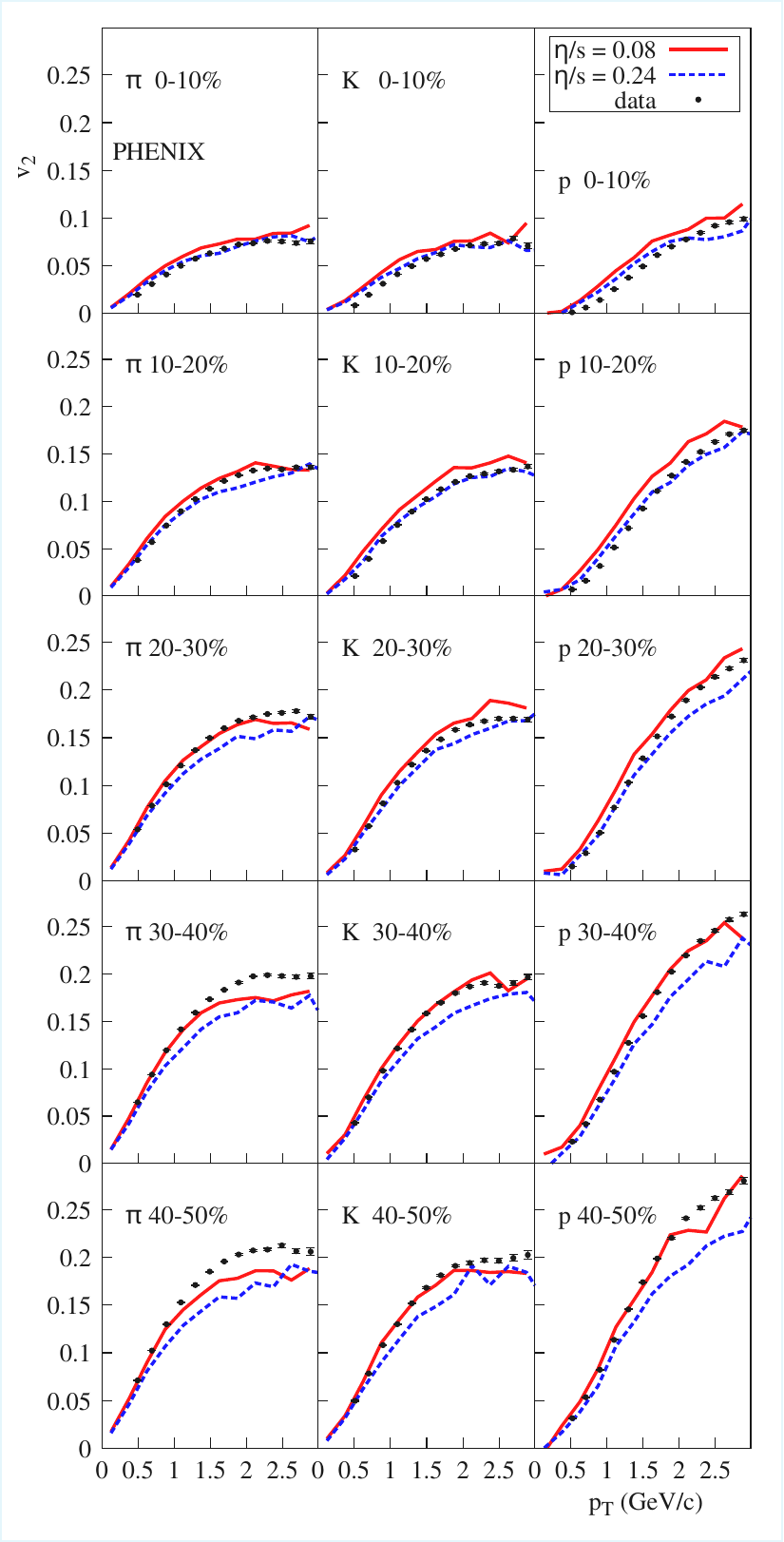}
\caption{Differential elliptic flow ($v_{2}$) of $\pi$, $K$, and $p$, as a function of transverse momentum $p_{T}$, at mid-rapidity, for Au+Au collisions at $\sqrt{s_{NN}}=200$ GeV, for different centrality bins.  The solid red curves refer to $\eta/s$ of 0.08, the blue dashed ones to 0.24. The black dots show the PHENIX experimental data \cite{PHENIX:2014uik}. }
\label{visc3}
\end{figure}

\begin{figure}
\centering
\includegraphics[width=\columnwidth]{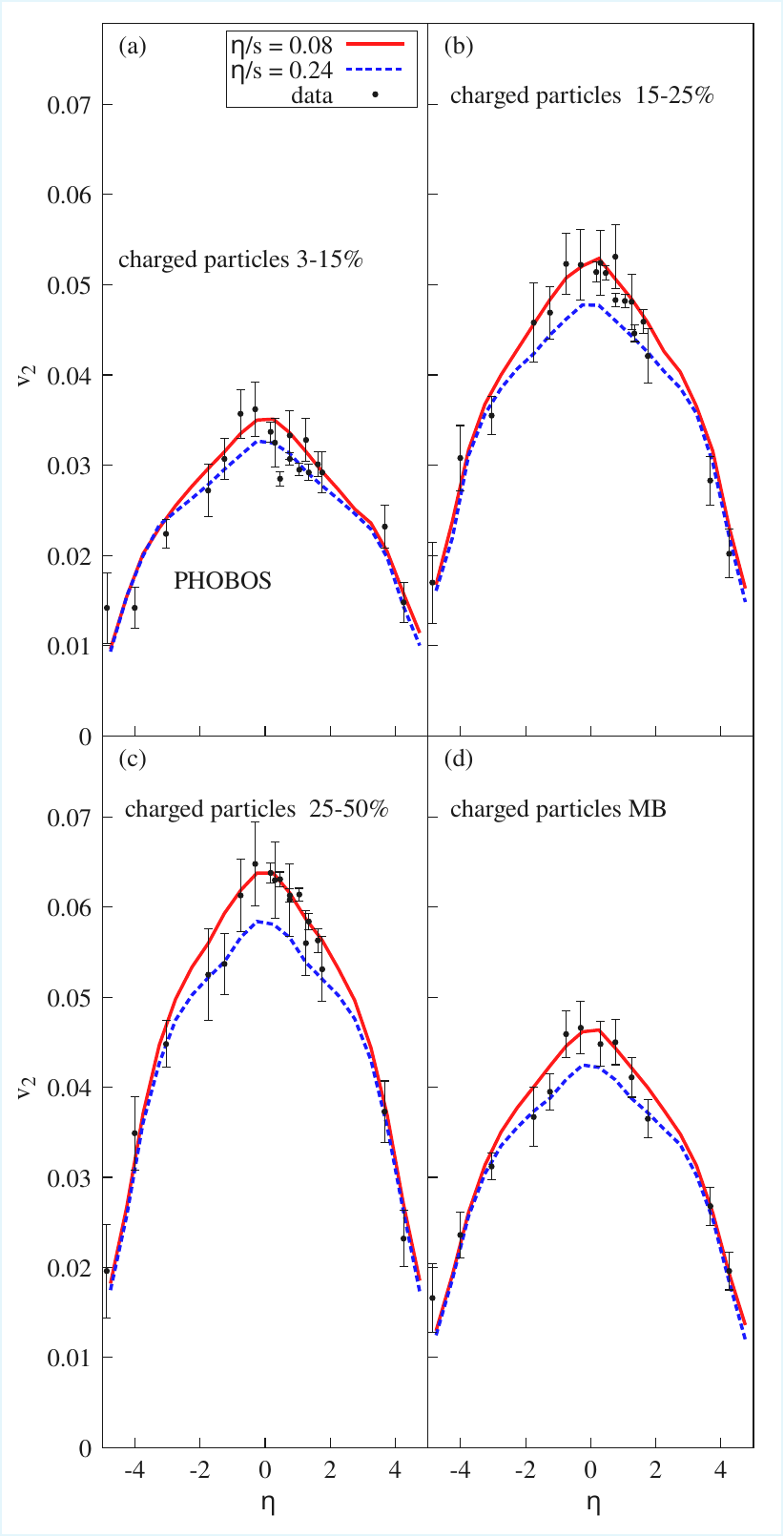}
\caption{Elliptic flow as a function of pseudorapidity ($v_{2}(\eta)$) for charged hadrons, in Au+Au collisions at $\sqrt{s_{NN}}=200$ GeV, for different centrality classes, i.e., (a): 3-15$\%$,  (b): 15-25$\%$, (c): 25-50$\%$, and (d): minimum bias (MB).  The solid red curves refer to $\eta/s$ of 0.08, the dashed blue ones to 0.24. The black dots show the PHOBOS experimental data \cite{PHOBOS:2004vcu}.}
\label{visc4}
\end{figure}

As already discussed earlier, the core-corona procedure is a crucial element in the EPOS picture, in particular the role of the core, expanding as a fluid, developing a strong flow. This has a very strong effect on spectra, as already discussed, but also on the $v_n$ distributions. We will discuss in the following the individuel contributions of core and corona for  $v_2$ and show results in Fig.  \ref{v2pt-coco}.
The other element to be discussed is the dependence of results on the shear viscosity, which will be studied as well in Figs.  \ref{visc3} and \ref{visc4}.


Concerning the core-corona contributions, we will distinguish again (A) the "core+corona" contribution, being the full simulation   (parallel scatterings, core-corona separation, hydrodynamic evolution, and microcanonical hadronization) but without hadronic rescattering;  (B) the "core" contribution being the same  as (A), but considering only core particles; (C) the "corona" contribution, being the same as (A), but considering only corona particles;  (D) the "full" EPOS4 scheme, namely as (A), but in addition hadronic rescattering. 
In Fig. \ref{v2pt-coco}, we show differential elliptic flow ($v_{2}$) of $\pi$, $K$, $p$, and charged particles, as a function of transverse momentum $p_{T}$, at mid-rapidity, for Au+Au collisions at $\sqrt{s_{NN}}=200$ GeV, for central (0-5$\%$) and mid-peripheral (40-50$\%$) collisions, for EPOS simulations. Thick red, dotted blue, green, and dashed yellow colors indicate the full (EPOS+hydro+UrQMD), corona, core, and core+corona results, while the black dots show the PHENIX experimental data \cite{PHENIX:2014uik}, respectively.

Naively one would expect zero elliptical flow for the corona part (dotted blue curves). This is actually what we get doing a simulation without hydro, where particle production comes from string decay, being perfectly symmetric.  Here, the values obtained from the simulation are very small, but not exactly zero. This is due to the fact that the core-corona procedure itself may introduce an azimuthal asymmetry. But the effect is small.

Looking at the $v_2$ curve for the core (green curve), it increases continuously, overshooting finally the data. This is in particular visible for the intermediate centralities (40-50\%) where the $v_2$ values are largest.  

Now let us look at the dashed yellow curves, the  core+corona contribution. We have seen that corona alone gives a very small $v_2$ result, therefore when considering both, particles from  corona and core, the former bring the $v_2$ of the core particles down. So the dashed yellow curves are somewhat below the green ones.  Interesting is the fact that the yellow core+corona curves even bend down at high  $p_{T}$, similar as the data. This "bending down" is a typical core-corona effect in the EPOS framework.  

Finally, we come to the thick red "full" curve, representing the final result, based on all particles. The difference between the yellow core+corona curve and the red one is the fact that the hadronic cascade is considered in the latter. So the effect of the hadronic cascade is  to increase $v_2$, more or less pronounced depending on the particle species.

So we see that also the $v_2$ curves (and more generally $v_n$) are strongly affected by the core-corona mechanism.

In Figs. \ref{visc3} and \ref{visc4}, we consider the $p_T$ and the $\eta$ distributions of $v_2$, compared to data, already seen earlier. But here we compare simulations for  $\eta/s$ of 0.08 (solid red curves) and those for $\eta/s$ of 0.24 (dashed blue curves). The difference is small. It should be said that EPOS employs "a rapid hadronization" according to microcanonic decay, and not a Cooper-Frye method with "viscous corrections". So in EPOS, the effect of viscosity is only due to the modification of the fluid dynamics, and not related to the hadronization.

\subsection{Messages from the flow harmonics comparison}

Considering the  $p_T$ dependence,  we observe that EPOS provides the best description  of $v_2$ for $\pi$, $K$, and $p$. We have seen that the hight $p_T$ behavior in EPOS is strongly affected by the core, representing hydrodynamic flow. The final phase (hadronic cascade) also matters, but crucial is the flow of the core, which develops very early.
EPOSir+PHSDe  and PHSD deviate from data with increasing $p_T$. 
This is in line with the time evolution of momentum eccentricity $\epsilon_P$ in Fig. \ref{eccp-compare}, where EPOS shows a stronger increase than EPOSir+PHSDe  and PHSD.
The fact that the flow harmonics $v_2(p_T)$ from EPOSir+PHSDe are closer to PHSD, not to EPOS, supports our conclusions from the previous section that the dynamical evolution is more important than the initial conditions.

Concerning the pseudorapidity dependence, we note that the ability of PHSD to describe well the flow $v_n$ data at midrapidity and at low $p_T \le 1$ GeV/c has been reported in our early studies \cite{Bratkovskaya:2011wp, Konchakovski:2011qa, Konchakovski:2014fya, Soloveva:2020ozg} and is related to the dominant role of the QGP at low $p_T$ at midrapidity. Since the QGP dynamics is well kept in PHSD by the DQPM model which follows the lQCD EoS, the PHSD  reproduces the quark number scaling \cite{Bratkovskaya:2011wp} and
provides a good description of the bulk and flow observables of final hadrons. 

The high $p_T$ part  is more sensitive to 
the spatial fluctuations in the initial conditions and the consequential 
strong radial flow, based on the core-corona separation, which is absent in the PHSD and is the key conceptual idea of the EPOS model. As has been discussed already several times, this core-corona separation in connection with the use of hydrodynamics --  which assumes a full early thermalization of the matter -- leads to the successful transformation of initial spatial anisotropy to final state momentum asymmetry. This is not observed in EPOSir+PHSDe.

\section{Summary}

In this study, we have investigated the relative impact of initial conditions and the dynamical evolution on the final observables in relativistic heavy-ion collisions. To achieve this goal, we developed a novel EPOSir+PHSDe framework, which utilizes EPOS initial conditions (EPOSi), complemented   by a rope mechanism (EPOSir),  followed by the non-equilibrium evolution within the PHSD approach (PHSDe).  
For comparison, we consider three models: 
\begin{enumerate}

\item  {\bf EPOS} employs instantaneous (parallel) multiple scatterings in an S-matrix approach, followed by a core-corona separation,  a hydrodynamic evolution of the quark-gluon plasma phase,  and hadronic rescatterings using UrQMD.  
\item  {\bf PHSD} initiates with primary high-energy independent nucleon-nucleon scattering via the LUND string model and evolves in time using fully microscopic transport dynamics for both partonic and hadronic matter.  
\item  {\bf EPOSir+PHSDe} combines the EPOS initial conditions with the PHSD dynamical evolution.  
\end{enumerate}

By comparing EPOSir+PHSDe with EPOS, we analyze the impact of different evolution models while keeping the initial conditions identical. Conversely, by comparing EPOSir+PHSDe with PHSD, we isolate the effects of different initial conditions for the same evolution model. This approach allows us to disentangle the influence of initial conditions from that of the dynamical evolution.  

For model validation, we focus on Au+Au collisions at an invariant energy of $\sqrt{s_{NN}} = 200$ GeV, where the QGP volume is substantial. We examine bulk matter observables and anisotropic flow to assess the role of initial conditions and the dynamical evolution in heavy-ion collisions.  
Our findings can be summarized as follows:

\begin{itemize}
\item 
We found that using  EPOSi leads to a strong overestimation of the final hadron yield when employing PHSDe. This "overproduction" is needed when using hydro, since the latter will reduce particle production because a large part of the energy is converted into flow and into work. The  introduction of the ropes, i.e. the method "EPOSi+ropes" (EPOSir in short),  decreased the number of prehadrons.  
Thus, the ropes allow to convert the initial large mass production in the Pomeron picture in EPOS to kinetic energy degrees-of-freedom.

\item 
By studying  the radial expansions via momentum eccentricities $\epsilon_P$, we find that the system in EPOSir+PHSDe and PHSD expands more slowly in transverse direction and $\epsilon_P$ reaches saturation compared to the EPOS momentum eccentricity, which grows monotonically in time. 
Moreover, the EPOS $\epsilon_P$  increases very strongly (with large fluctuations) at very early times. In EPOSir+PHSDe,  $\epsilon_P$ even decreases right after the start time, showing much less fluctuations compared to EPOS. The average  $\left<\epsilon_P\right>$ saturates at approximately the same value as in PHSD. 
This highlights the dominant influence of the evolution dynamics over the initial conditions.
The difference in modeling the evolution between EPOS and PHSD leads to distinct behaviors in $\epsilon_P$. In EPOS, the monotonic increase in momentum eccentricities is driven by its hydrodynamic description of the QGP. 
Conversely, in PHSD the QGP dynamics is modeled using the lQCD equation of state (EoS) and described in terms of off-shell massive partons within the DQPM framework. 

\item By studying the transverse momentum/mass spectra as well as rapidity distributions of hadrons, we find that EPOS accurately reproduces data for the charged particle spectra at all $p_T$  including intermediate and high $p_T$, while PHSD and EPOSir+PHSDe show more soft spectra, and in spite that both models describe well the low part of the $p_T$ spectra, they  deviate from the data at intermediate and high $p_T$. 

\item Similar observations hold for the $p_T$ behavior of the flow harmonics $v_2(p_T)$: a very good description of $v_2$ for all $p_T$  by EPOS and deviation from the data for EPOSir+PHSDe and PHSD with increasing $p_T$. On the other hand, PHSD provides a good description of the pseudorapidity distribution of $v_2(\eta)$  which is dominated by low momentum hadrons.
\end{itemize}

The observation that EPOS and EPOSir+PHSDe, despite starting from similar initial conditions but following different dynamical evolutions, yield different results for hadron spectra and flow harmonics $v_n(p_T)$, and that EPOSir+PHSDe and PHSD, which begin with entirely different initial conditions but follow a similar dynamical evolution, produce approximately the same hadron spectra and $v_n$, highlights the dominant influence of the dynamical evolution on these observables over the initial conditions. The dynamical evolution finally defines how the initial conditions are evolved in time.

We can highlight the most important aspects in modeling of the dynamical evolution in the interplay with the initial conditions using the EPOS, PHSD and EPOSir+PHSDe as follows:

\begin{itemize}
\item 
The ability of EPOS for a very good description of the $p_T$-dependence of bulk and flow observables can be attributed to
\begin{itemize}
\item
the modeling of the initial stage using instantaneous (parallel) multiple scatterings, with a strongly fluctuating space anisotropy, 
\item
the separation of the produced particles into core and corona, where the former is  converted to a hydrodynamical description (which assumes local equilibration) at an early time, which creates very quickly flow and large (strongly fluctuating) momentum anisotropies,
\item
where the hydrodynamic evolution also converts a large fraction of the initial system energy into kinetic energy of expanding matter.
\end{itemize}
\item 
In PHSD the good description of low $p_T$ and underestimation of the high and intermediate  $p_T$ hadron spectra and flow $v_n$  can be related (in comparison to EPOS) to:
\begin{itemize}
\item 
building of the initial phase  by independent sequences of high energy $NN$ scatterings (using the LUND string model)  which leads to  more smooth initial conditions than in EPOS, 
\item
being a fully microscopic non-equilibrium model, PHSD propagates hadronic and partonic degrees of freedom explicitly. On one side it allows to avoid discontinuities (e.g. a jump in entropy and energy density) attributed to the transition from initial conditions to a hydrodynamical description as well as to the hadronic phase after the hadronization procedure.  On the other side, it requires the interpretation of the EoS in terms of properties of degrees of freedom and their interactions: the QGP phase is modeled by interacting off-shell quasipartcles within the DQPM model which reproduces the lQCD thermodynamics near equilibrium and leads to a large and temperature dependent  $\eta/s(T)$, which does not lead to a very early thermalization as assumed in the EPOS hydrodynamical expansion.
\end{itemize}
\item 
The underestimation by EPOSir+PHSDe (similar to the PHSD) of the intermediate and high $p_T$ range of hadron spectra and flow coefficients indicate that the initial space anisotropy -  from the Pomeron breaking as in EPOS - is less  developing into flow and anisotropic flow in the microscopic non-equilibrium dynamics of PHSD during the partonic and hadronic phases. All that leads to more soft spectra of  EPOSir+PHSDe, which  deviate form the experimental data with increasing $p_T$.
\end{itemize}

Thus, by developing EPOSir+PHSDe we obtain an interesting tool to investigate the different aspects of heavy-ion dynamics in order to optimize their modeling based on a solid theoretical ground.

\begin{acknowledgments}
The authors thank J. Aichelin  for useful discussions and valuable suggestions and to I. Karpenko for the guide to the details of hydrodinamical models.
We are grateful to W. Cassing for a careful reading of the manuscript and important comments. 
We also acknowledge the support by the Deutsche Forschungsgemeinschaft (DFG) through the grant CRC-TR 211 "Strong-interaction matter under extreme conditions" (Project number 315477589 - TRR 211). 
\end{acknowledgments}

\bibliography{references}

\end{document}